\documentclass{emulateapj}
\usepackage{amsmath}
\usepackage{rotating}
\usepackage{epsf}
\usepackage [english]{babel}
\usepackage [autostyle, english = american]{csquotes}
\usepackage{chngcntr}
\MakeOuterQuote{"}
\pdfoutput=1 
\usepackage{graphicx}
\usepackage{verbatim}
\usepackage{color}
\usepackage{longtable}
\newcommand{\myemail}{bajoshi@asu.edu}
\newcommand{\red}{\textcolor{red}}
\newcommand{\kms}{km~s$^{-1}$}
\shorttitle{}
\shortauthors{Joshi et al.}

\received{December 18, 2018}
\accepted{May 11, 2019}

\begin{document}

\title{Evidence for shock-heated gas in the Taffy Galaxies and Bridge from Optical Emission-Line IFU spectroscopy}

\author{Bhavin A.\ Joshi\altaffilmark{1,2}, Philip N.\ Appleton\altaffilmark{2}, Guillermo A.\ Blanc\altaffilmark{3,4}, Pierre Guillard\altaffilmark{5,6}, Jeffrey Rich\altaffilmark{3}, Curtis Struck\altaffilmark{7}, Emily E.\ Freeland\altaffilmark{8}, Bradley W.\ Peterson\altaffilmark{9}, George Helou\altaffilmark{2}, Katherine Alatalo\altaffilmark{3,10}}

\altaffiltext{1}{School of Earth and Space Exploration, Arizona State University, Tempe, AZ 85287, USA}
\email{\myemail}
\altaffiltext{2}{IPAC, Caltech, MC 100-22, 770 S Wilson Ave., Pasadena, CA 91125, USA}
\email{apple@ipac.caltech.edu}
\altaffiltext{3}{The Observatories of the Carnegie Institution for Science, 813 Santa Barbara Street, Pasadena, CA 91101, USA}
\altaffiltext{4}{Departamento de Astronom\'ia, Universidad de Chile, Casilla 36-D, Santiago, Chile}
\altaffiltext{5}{Sorbonne Universit\'e, CNRS, UMR 7095, Institut d'Astrophysique de Paris, 98bis bd Arago, 75014 Paris, France}
\altaffiltext{6}{Institut Universitaire de France, Minist\`ere de l'Education Nationale, de l'Enseignement Sup\'erieur et de la Recherche, 1 rue Descartes, 75231 Paris Cedex 05, France}
\altaffiltext{7}{Department of Physics and Astronomy, Iowa State University, Ames, IA 50011, USA}
\altaffiltext{8}{Department of Astronomy, Oskar Klein Centre, Stockholm University, SE-10691 Stockholm, Sweden}
\altaffiltext{9}{Hastings College, 710 N.\ Turner Ave., Hastings, NE 68901, USA}
\altaffiltext{10}{Space Telescope Science Institute, 3700 San Martin Drive, Baltimore, MD 21218, USA}

\begin{abstract}
We present optical IFU observations of the Taffy system (UGC 12914/15); named for the radio emission that stretches between the two galaxies. Given that these gas rich galaxies are believed to have recently collided head-on, the pair exhibits a surprisingly normal total (sub-LIRG) IR luminosity ($\mathrm{L_{FIR} \sim 4.5 \times 10^{10}}$ L$_\odot$). Previous observations have demonstrated that a large quantity of molecular and neutral gas have been drawn out of the galaxies into a massive multi-phase bridge. We present, for the first time, spatially resolved spectroscopy of the ionized gas in the system. The results show that the ionized gas is highly disturbed kinematically, with gas spread in two main filaments between the two galaxies. The line profiles exhibit widespread double components in both the bridge and parts of the disks of the galaxies. We investigate the spatial distribution of the excitation properties of the ionized gas using emission-line diagnostic diagrams, and conclude that large quantities (up to 40$\%$) of the emission from the entire system is consistent with gas heated in $\sim$200 \kms\ shocks. While the shocked gas is mainly associated with the bridge, there is a significant amount of shocked gas associated with both galaxies. Confirming other multi-wavelength indicators, the results suggest that the effects of shocks and turbulence can continue to be felt in a high-speed galaxy collision long after the collision has occurred. The persistence of shocks in the Taffy system may explain the relatively low current star formation rates in the system as a whole.
\end{abstract}


\section{Introduction}
It is now generally accepted that collisions and mergers between gas-rich galaxies often generate intense star-formation activity and associated strong infrared emission \citep[e.g.,][]{Joseph1985, Soifer1987, Soifer1991}. Ultra Luminous Infrared Galaxies, ULIRGs (L$_\mathrm{IR}$ $>$ 10$^{12}$L$_{\odot}$) and LIRGS (10$^{12}$ $>$ L$_\mathrm{IR}$/L$_{\odot} >$ 10$^{11}$) frequently involve mergers or interactions of gas-rich galaxy pairs, with the likelihood of them being associated with a major merger increasing with infrared (IR) luminosity \citep{Armus1987, Sanders1988a, Sanders1988b, Sanders1996, Elbaz2002, Armus2009}. While it is clear that major mergers play an important role in generating high IR luminosities in the local universe, their role at higher redshift is still being explored. 

Shocks and turbulence potentially play a role in changing the conditions of the gas in collisional galaxies, not always leading to enhancements in star formation. In the local Hickson Compact Groups, \citet{Alatalo2015} found evidence that multiple collisions can quench or significantly suppress star formation in some systems where turbulence and shocks are present \citep[see also][]{Lisenfeld2017}. These galaxies had previously been found to contain large volumes of warm molecular hydrogen that emit their energy mainly in the mid-IR, and were believed to be shock-heated \citep{Cluver2013}. An extreme example is found in the Stephan's Quintet system. Here, a large filament of molecular gas is found in the intergalactic medium in which a large fraction of the gas is warm and in a shock-heated phase \citep{Appleton2006, Guillard2009, Cluver2010, Appleton2017}. Shocks, though hard to detect in LIRGs and ULIRGs because of the dominant effects of star formation on optical emission-line diagnostics, are being increasingly detected with the advent of spatially resolved optical integral field unit (IFU) spectroscopy \citep{Rich2011, Rich2014, Rich2015}. How large-scale shocks and turbulence affect the star formation in such galaxies, and how important this process is in higher-redshift systems is currently unknown. 

An interesting example of an ongoing major merger that may be caught in a  highly disturbed state is the Taffy galaxy pair UGC 12915/4 (hereafter Taffy-N and Taffy-S for simplicity).  Despite having recently undergone a strong head-on collision, the Taffy system appears surprisingly normal in its IR properties, with a total L$_\mathrm{IR}$ = 4.5 $\times$ 10$^{10}$ L$_{\odot}$ summed over the whole system based on multi-wavelength {\it Spitzer} and {\it Herschel} SED photometric fitting \citet{Appleton2015}; \citep[see also][]{Jarrett1999, Sanders2003}. The reason that the system is so normal in the IR, despite its recent violent history, is not known. It may be that we are catching the Taffy system in a peculiar moment where most of its gas is so disturbed that it cannot yet generate significant star formation. If so, studying the conditions of the gas in between the galaxies (referred to as the bridge) may well yield interesting insight into how shocks and turbulence can inhibit star formation in violently colliding galaxies. 

The Taffy galaxies were named for the discovery of a bridge of radio continuum emission, stretching, like salt-water taffy (candy), between the galaxies \citep{Condon1993}.  Evidence suggests that the two galaxies collided 25-30 Myr ago, allowing their stellar components to pass through each other, but stripping $\sim$ 7 x 10$^9$ M$_{\odot}$ of molecular and atomic gas into a bridge between them \citep{Braine2003, Gao2003, Zhu2007}. There is more gas in the bridge than in the two galaxies combined.  

The bridge appears to be strongly disturbed (and probably turbulent), based on kinematically-broad CO line studies of the bridge, and strong mid-IR H$_2$ emission and [CII]157.7$\mu$m lines suggestive of shocks \citep{Peterson2012, Peterson2018}. Despite its high gas mass, the average star formation rate (SFR) in the entire bridge through SED fitting is quite low, $\sim$0.45 M$_{\odot}$ yr$^{-1}$, excluding the prominent extragalactic HII region seen south-west of UGC 12915, which was separately found to have a SFR of 0.24 M$_{\odot}$ yr$^{-1}$ \citep{Appleton2015}. Numerical models of such a head-on collision between two gas rich galaxies \citep[e.g.,][]{Struck1997} and a detailed model of the Taffy system \citep{Vollmer2012} provide strong support for the idea that the gas left behind in the center of mass frame of the collision would be highly turbulent, and that some would be strongly shock heated. \citet{Appleton2015} detected faint extended soft X-ray emission, and several compact point X-ray sources in the bridge, the former being consistent with shock-heated gas that has not had time to completely cool since the collision occurred. Finally, \citet{Lisenfeld2010} concluded that the radio emission in the bridge could be explained in terms of cosmic rays accelerated in magnetic fields compressed in shocks.

Although the Taffy galaxies have been studied quite extensively at longer wavelengths \citep{Condon1993, Jarrett1999}, very little work has been done at visible or near-IR wavelengths. \citet{Bushouse1987} presented early digital video camera observations which showed H$\alpha$ emission from the inner disks of both galaxies and emission from the extragalactic HII regions in the bridge. Pa$\alpha$ observations from the ground were also made by \citet{Komugi2012}. The galaxies show strong disturbances in their optical structure, including rings and loops, and the possible recent onset of star formation in the bridge, including at least one prominent extragalactic HII region, and fainter clusters--some of which are seen in archival NICMOS observations from HST (Appleton et al.\ in preparation).

This paper represents the first major study of the ionized gas phase in the Taffy system and bridge. We provide, for the first time, a detailed exploration of both the kinematics and excitation properties of the optical emission line gas in the Taffy system. The paper is organized as follows: \S\ref{sec:data_methods} describes the observations and methods used in the paper. We describe the fitting process used on the double line profiles and the gas kinematics through H$\alpha$ channel maps and velocity field moment maps in \S\ref{sec:double_profiles} and \S\ref{sec:kinematics}, respectively. The effects of dust extinction on the measured line fluxes are discussed in \S\ref{sec:dust}. We describe the results from our line diagnostic diagrams in \S\ref{sec:bpt_results}. In \S\ref{sec:frac_from_sf}, we discuss our results on the properties of the ionized gas and its excitation mechanisms through the use of emission-line ratio diagnostic diagrams and comparison with shock models. In \S\ref{sec:frac_from_sf} and \S\ref{sec:sfr} we discuss our estimates of the ionized gas fraction from star formation and the SFR in the system. We discuss the evidence for a post-starburst population in the underlying starlight in \S\ref{sec:hbeta_ew_results}. In \S\ref{sec:conclusions} we present our conclusions. We assume a comoving distance to the galaxies of 62 Mpc based on a  mean heliocentric velocity for the system of 4350 \kms, and a Hubble constant of 70 \kms Mpc$^{-1}$.

\section{Observations, Data Reduction, and Analysis Methods}
\label{sec:data_methods}

The IFU data presented in this work were obtained with the VIRUS-P Spectrograph at McDonald Observatory \citep{Hill2008,Blanc2010}. VIRUS-P is the Visible Integral-field Replicable Unit Spectrograph prototype (now called the George and Cynthia Mitchell Spectrograph, GCMS)  mounted on the 2.7 m Harlan J.\ Smith telescope. The IFU has 246 fibers (each fiber has an angular diameter of $4\farcs16$ on the sky) with a $\frac{1}{3}$ filling factor. We used several cycles of a 3-point dither pattern to completely cover the 2.8 sq.\ arcminute field of view.  We used the gratings VP2 and VP4 for our blue and red channel spectra respectively. VP2 and VP4 have a spectral resolution of 1.6 and 1.5~\AA\, and covered a range of 4700--5350~\AA\ and 6200--6850~\AA, respectively. This spectral resolution corresponds to a velocity resolution of $\sim$100~\kms and $\sim$70~\kms at the wavelengths of H$\beta$ in VP2 and H$\alpha$ in VP4, respectively.

We made observations of the Taffy galaxies on 2012 Jan 31 and Feb 2 (blue spectrometer) and 2012 Feb 01 (red spectrometer) with a total exposure time per dither position of 2200s  (blue) and 1200s (red). Conditions were photometrically good at the time of the observations with moderate seeing of 1.8-2.5 arcseconds (less than the diameter of a fiber).  

These data were processed using the VACCINE pipeline which identified and traced each fiber on the CCD chip, and performed bias, flat-field and wavelength calibration (based on lamp spectra) on a fiber-by-fiber basis for the science frames. VACCINE is a Fortran-based reduction package developed for the HETDEX Pilot Survey \citep{Adams2011} and the VENGA project \citep{Blanc2010}. Cosmic-ray removal was then performed using the IDL routine LA-Cosmic \citep{vanDokkum2001}.

\subsection{Flux Calibration and Cube Building}
\label{sec:data_reduction}

The flux calibration and pointing refinements and final cube construction of these data was performed in three steps following the methods described by \citet{Blanc2013}: i)  Relative spectrophotometric calibration was performed which is applied to all the fibers. Observations were made in a 6-point dither pattern (including several fibers) of the standard star  Hz 15 (HIP 21776). An algorithm was used to solve for the position of the star on the fibers and determine the spectrophotometric transformation from native (ADU) units across the spectrum to units of erg s$^{-1}$ cm$^{-2}$ \AA$^{-1}$. The results were applied to all the fibers irrespective of throughput. This step resulted in a relative flux uncertainty across the band of $\sim$8$\%$;  
ii) astrometry and absolute flux calibration, using a bootstrapping method, was used to effectively cross-correlate a reconstructed image of the galaxy derived by integrating the light from each fiber, with a calibrated images of Taffy from the SDSS \citep{York2000} in the g- and r-band, suitably convolved to the resolution of the VIRUS-P fiber system. This helped refine the astrometry, and the assembly of the final cube from the individual observations of each field. The cross-correlation also allowed the spectrum in each fiber to be absolutely scaled to the SDSS band in question. The details of this procedure are given in \citet{Blanc2013}. Tests performed in that paper show that the absolute spectrophotometric flux calibration has a typical accuracy of 15-30$\%$, after taking into account the uncertainties in SDSS calibration and the VIRUS-P relative spectrophotometric accuracy;  
iii) a final flux-calibrated 3-d spectral cube was created by combining all the various observational pointing frames into a single interpolated cube with resulting 2 x 2 arcsec$^2$ spaxels ($\sim$0.3 kpc arcsec$^{-1}$ based on the assumed distance of 62 Mpc). These processes were repeated for the red and blue channels, creating final flux-calibrated blue and red spectral cubes.

\subsection{Spectral Mapping, continuum and emission-line fitting}
\label{sec:data_processing}
The processing and extraction of astrophysical information from the data cubes was done using a combination of IRAF/PyRAF, IDL and Python routines. Before beginning our analysis we smoothed the data cubes spatially, but \emph{not} in the spectral direction, using a Gaussian kernel with a standard deviation of 1.47 pixels. This was done to boost the signal-to-noise ratio in areas that we were interested in; particularly the Taffy bridge region which has relatively low signal-to-noise compared to the galaxies. This spatial smoothing effectively reduces the noise in spectra from individual spaxels by a factor of $\sim$2. We used these spatially smoothed cubes for all of the analysis done in this work.

\begin{figure}
\centering
\includegraphics[width=0.48\textwidth]{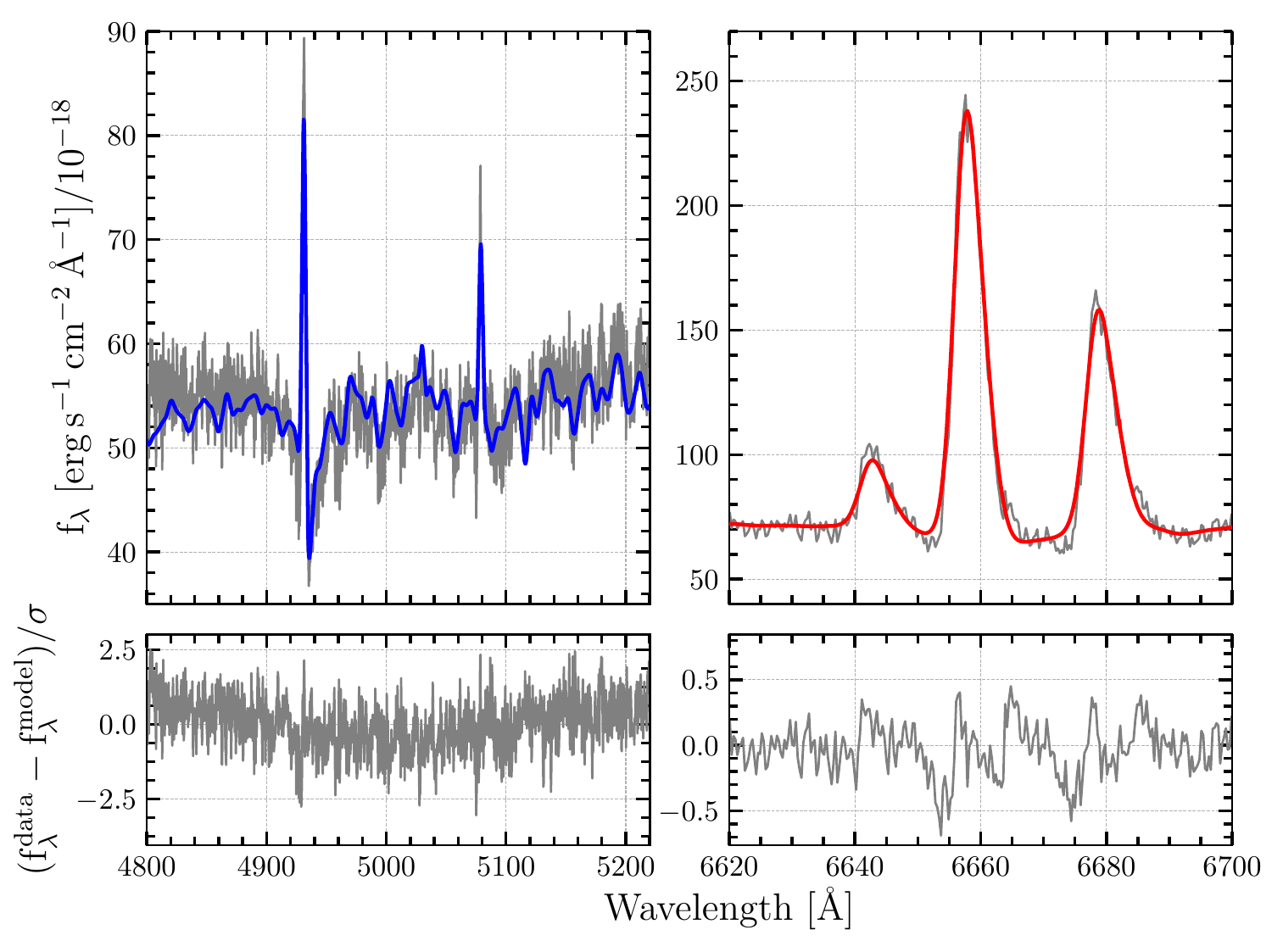}
\caption{Our LZIFU fitting results for a spaxel that has deep H$\beta$ absorption and H$\beta$ emission superimposed on the absorption trough. The top left and right panels show the blue and red data respectively, along with their model fits. The bottom panels show the corresponding residuals from the fitting. The gray line shows the raw data from the spaxel and the blue and red lines show the model fits to the respective channels. Note that the Figure does not show the full wavelength coverage of the data but instead is focused on showing the relevant absorption and emission features i.e. H$\beta$ and the [OIII]$\lambda\lambda$4959,5007 doublet in the blue channel and H$\alpha$ and its neighboring [NII]$\lambda\lambda$6548,6583 doublet lines in the red channel.}
\label{fig:lzifu_fit_hbeta_abs}
\end{figure}

\begin{figure}
\centering
\includegraphics[width=0.48\textwidth]{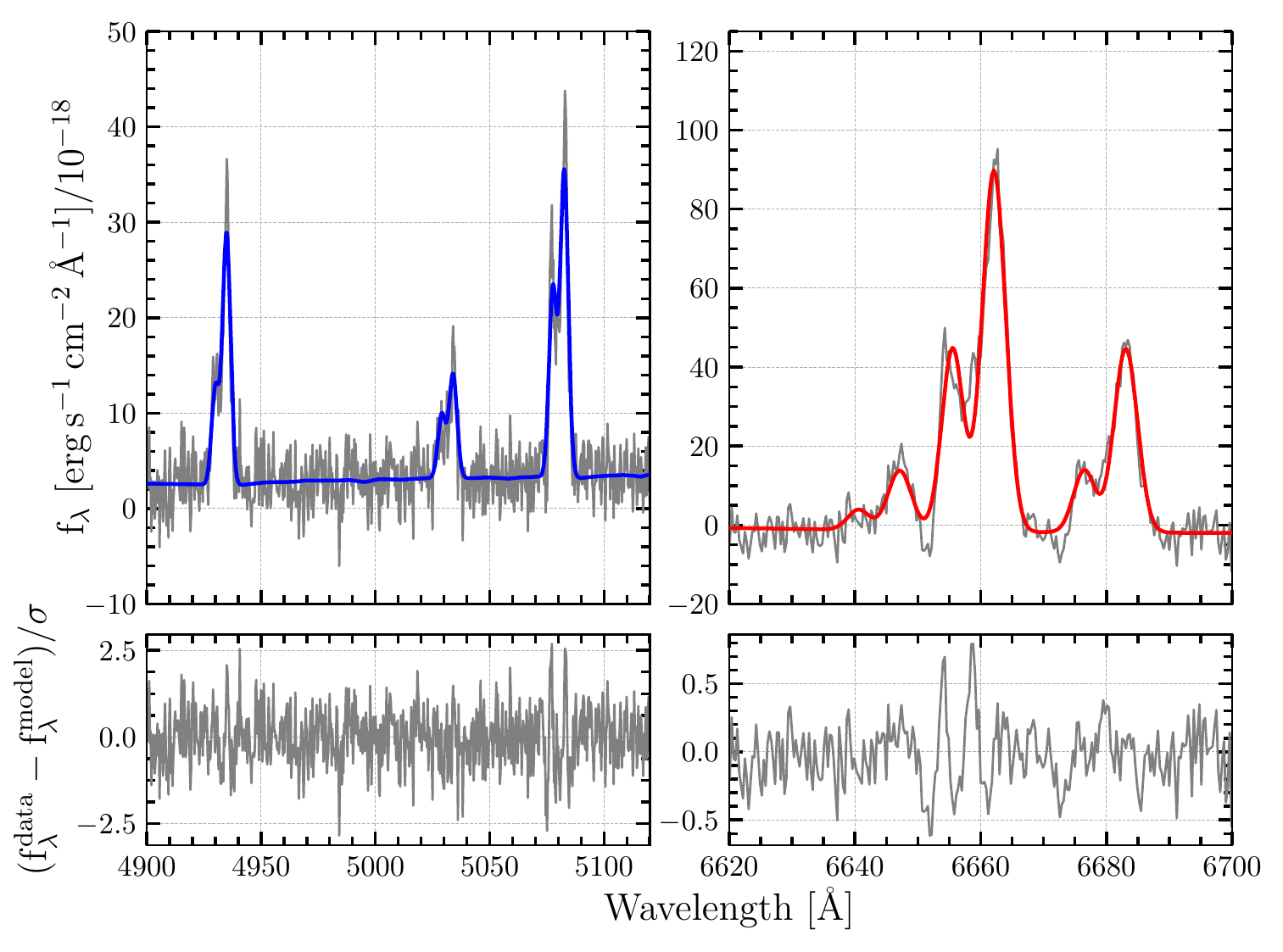}
\caption{Same as Figure \ref{fig:lzifu_fit_hbeta_abs} but now showing the LZIFU fitting results for another spaxel which displays distinct velocity components more clearly.}
\label{fig:lzifu_fit_vel_comp}
\end{figure}

For fitting each individual spaxel in the IFU data we used the IDL software toolkit LZIFU (LaZy-IFU) \citep{Ho2016}. LZIFU automates fitting multiple emission lines superimposed on a continuum for multiple spaxels in each channel and provides 2D maps of continuum and line fluxes, velocities and velocity dispersion. It is capable of fitting emission lines that are superimposed on deep absorption features and also emission lines with multiple velocity components. Figures \ref{fig:lzifu_fit_hbeta_abs} and \ref{fig:lzifu_fit_vel_comp} show our fitting results for two individual spaxels that show these features in their spectra. Figure \ref{fig:lzifu_fit_hbeta_abs} shows a spaxel that has strong H$\beta$ absorption and H$\beta$ emission superimposed on the absorption trough. This spaxel lies very close to the center of Taffy-N. This absorption must be accounted for with the continuum fitting to get accurate emission line fluxes as well as an accurate line profile. The emission lines also show evidence for double line  profiles in many positions across the system. As an example, Figure \ref{fig:lzifu_fit_vel_comp} shows a spaxel which lies close to the edge of the extragalactic HII region which clearly shows two separate velocity components.

LZIFU works by first fitting the continuum with a custom implementation of the PPXF code \citep{Capellari2004} within LZIFU and then fitting the emission lines after subtracting the continuum model. The continuum fitting uses a set of model templates that are fit to a combined blue+red channel spectrum. LZIFU also accounts for systematic errors in the models and non-stellar contributions to the continuum data by fitting a multiplicative polynomial simultaneously with the continuum models. The emission lines along with residuals from sky line subtraction are masked during the continuum fitting process. We also specified the following lines to be fit (and masked during continuum fitting) - H$\beta$, the [OIII]$\lambda\lambda$4959,5007 doublet, [OI]$\lambda$6300, [OI]$\lambda$6364, H$\alpha$, the [NII]$\lambda\lambda$6548,6583 doublet, and the [SII]$\lambda\lambda$6716,6731 doublet. The extinction corrected emission line fluxes for all the lines detected in different regions of the Taffy system (as defined in Figure \ref{fig:Fig3}) are tabulated in Table \ref{tab:flux_table}.

We ran LZIFU on the entire IFU data cube for the Taffy galaxies which is $58\times58$ and $59\times59$ spaxels with 2227 and 2350 spectral elements for the blue and red channels respectively. The process of fitting the full cube was not straightforward for several reasons relating to the peculiar kinematics of the Taffy system. The LZIFU software was designed to work best with a galaxy showing slowly-varying velocity centroids relatively close to the initial guess for the velocity of the system. In the Taffy system, the velocity range of the emission lines over the whole system was large, with the emission lines sometimes exhibiting complex behavior, in addition to occasionally being observed superimposed on deep Balmer absorption lines. Furthermore, in a large number of spaxels, we found multiple velocity components which did not always move together as a function of position.  As a result, a single set of initial guesses for the various starting parameters did not work for the whole cube, but had to be adjusted spatially to achieve good fits, especially in specific regions showing double line profiles. After an iterative process of fitting with different starting parameters tuned to particular regions,  we were able to get a consistent set of smoothly varying results across the whole system. We provide the most relevant LZIFU parameters and their starting guesses in Table \ref{tab:lzifu_params}.

\section{Results}
\subsection{Emission-line gas and H$\beta$ Absorption within the System}
\label{sec:double_profiles}
Previous observations of the Taffy system in the visible wavelength range had reported the detection of ionized gas within the galaxies and the extragalactic HII region \citep{Bushouse1986, Bushouse1987}. Because of the sensitivity of the VIRUS-P instrument to very faint diffuse emission, we report the presence of ionized gas throughout the Taffy system -- both within the galaxies, the so-called extragalactic HII region, and also \emph{in the bridge between the galaxies}.  We detect emission from many lines, including H$\alpha$, H$\beta$, the [OIII]$\lambda\lambda$4959,5007 doublet, the [NII]$\lambda\lambda$6548,6583 doublet, and the [SII]$\lambda\lambda$6716,6731 doublet lines (see Figures \ref{fig:lzifu_fit_hbeta_abs} and \ref{fig:lzifu_fit_vel_comp}). We also detect strong emission from the atomic oxygen line [OI]$\lambda$6300 and sometimes the weaker [OI]$\lambda$6364 line in the galaxies and the extragalactic HII region in the bridge. Although emission lines dominate in many of the locations across the system, in some cases H$\beta$ emission is observed superimposed on a broad absorption trough indicative of a post-starburst population (as seen in Figure \ref{fig:lzifu_fit_hbeta_abs}).

\begin{figure*}
\centering
\includegraphics[width=\textwidth]{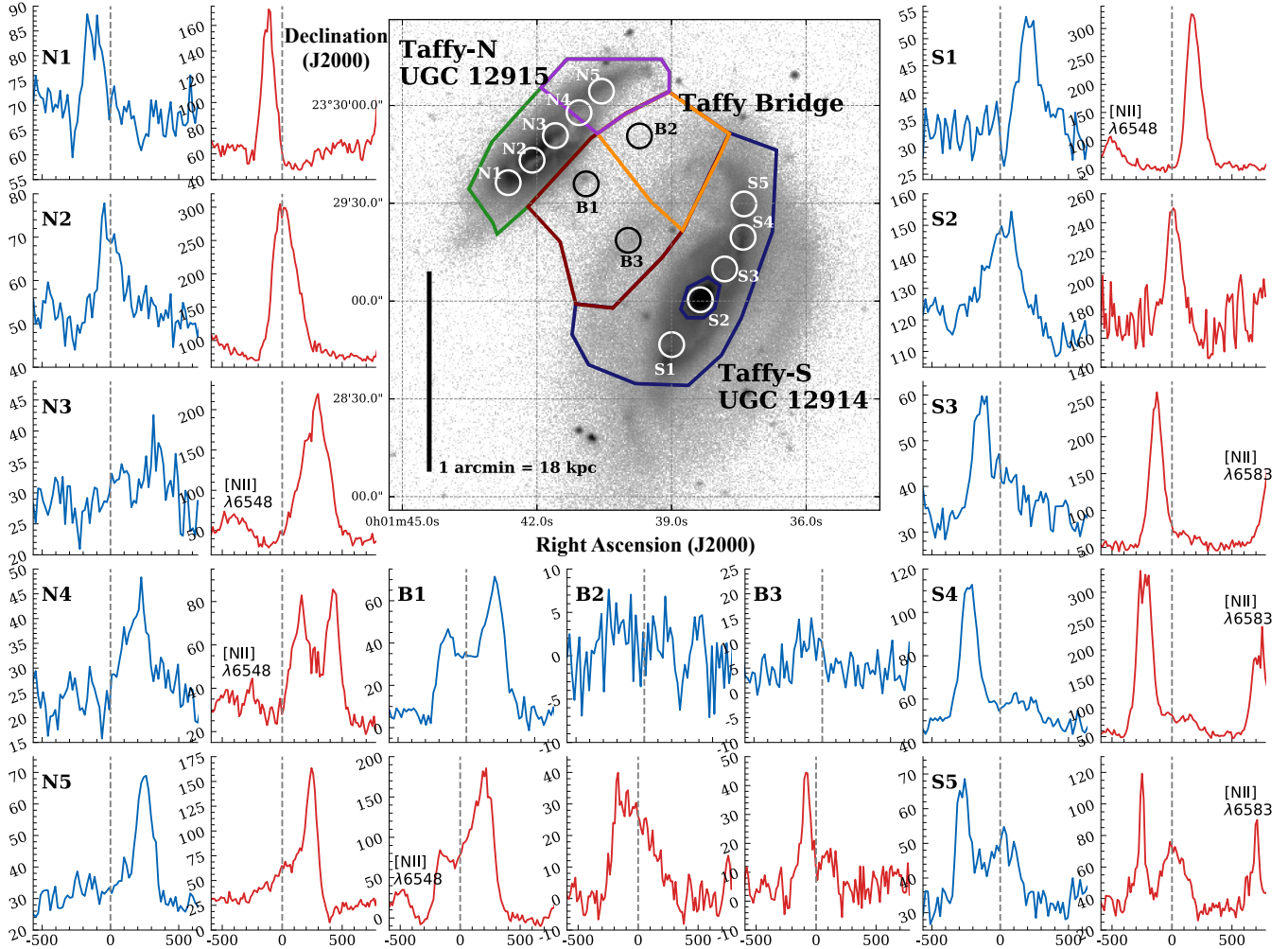}
\caption{Some examples of integrated VIRUS-P spectra from the Taffy system expanded to emphasize the kinematics. The galaxies UGC 12915/4 referred to as Taffy-N and Taffy-S respectively for clarity in the text are shown in this SDSS i-band image.   Spectra are shown extracted from the regions of the black and white colored circles.  The blue and red spectra correspond to the [OIII]5007 and H$\alpha$ lines. The flux axis is in units of $10^{-18}\, \mathrm{erg\, s^{-1}\, cm^{-2}\, \AA^{-1}}$.  The polygon regions refer to the regions that are color coded, using the same colors, in the later discussion of the emission-line diagnostic diagrams (with the exception of the western part of Taffy-N which is shown as green pluses in the line diagnostic diagrams). We denote Taffy-N, the western part of Taffy-N, Taffy-S, the eastern bridge, and the western bridge by green, magenta, blue, red, and orange colored polygons respectively. The nucleus of Taffy-S is also denoted by a small blue polygon. Our justification for selecting these regions is discussed in the text. The region shown here as B1 is centered on a faint extragalactic HII region discussed in the text. The gray dashed line denotes the systemic recessional velocity for the Taffy pair at 4350 km/s.}
\label{fig:Fig3}
\end{figure*}

In Figure \ref{fig:Fig3} we present some extracted spectra in several places in the system to provide an overview of the complexity of the kinematics in this system. The spectra show expanded views of the [OIII]$\lambda$5007 and H$\alpha$ lines along the major axis of both galaxies and a sampling of the bridge. As with the previous HI investigations of Taffy by \citet{Condon1993, Braine2003}, the ionized gas spectra along the major axis of Taffy-N (N1-N5) show clear rotation from low to high velocities as one proceeds northwards, whereas in Taffy-S (S1-S5), the rotation is also obvious, but in the opposite sense. This confirms the suggestion that the galaxies were counter-rotating when they collided. Many of the line profiles are complex, and contain multiple components. Of special note are the broad lines in the bridge (especially B1 and B2) as well as complex multi-component structures in the north-west of both galaxies (N4, N5, S4 and S5). The nucleus of Taffy-S (S2) also shows very broad strong [OIII] emission, and weaker broad wings in H$\alpha$ (especially when a correction is made for H$\alpha$ absorption--see \S4.2). We also show polygons marking additional regions of interest on the SDSS image in Figure \ref{fig:Fig3}. These polygons show regions which are investigated in the emission-line diagnostic diagrams in \S\ref{sec:bpt_results}. In the emission-line diagnostic diagrams, we investigate the excitation mechanisms for the western regions of Taffy-N and the bridge separately because these regions exhibit peculiar kinematics distinct from the rest of Taffy-N and the bridge (see the detailed discussion in \S\ref{sec:vel_map} and \S\ref{sec:methods_vel_comp}).

\begin{figure*}
\centering
\includegraphics[width=0.97\textwidth]{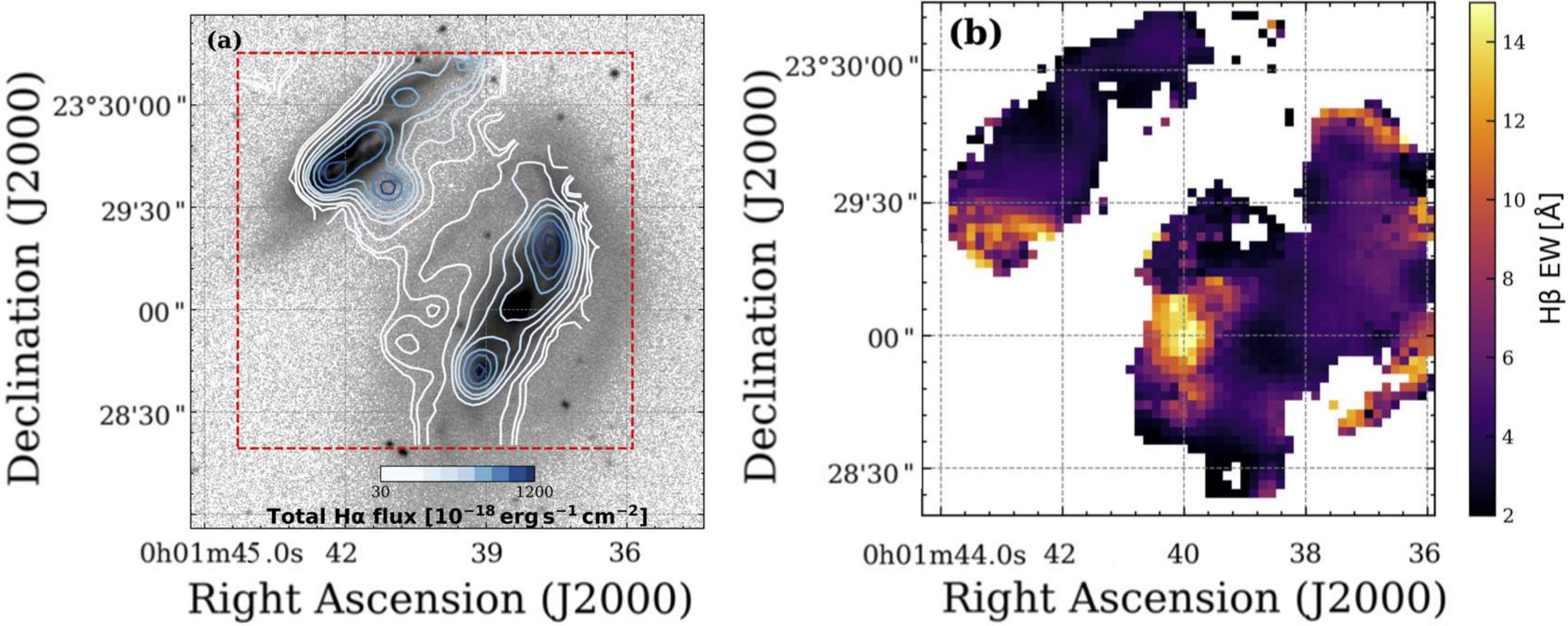}
\caption{(a) Integrated H$\alpha$ emission contours from the IFU data overlaid on an SDSS i-band image. The contour levels are 30, 50, 100, 200, 300, 400, 600, 800, 1000, and 1200, in units of $10^{-18}$ ${\rm erg\, s^{-1}\, cm^{-2}}$. The lowest contour level corresponds to an approximately 2$\sigma$ detection. The red rectangle shows the IFU coverage. The H$\alpha$ emission was summed over the two velocity components in spaxels where there were double profiles. (b) The measured H$\beta$ absorption-line equivalent width (EW) in Angstroms across the Taffy system based on the ppxf fitting of the absorption lines and continuum.}
\label{fig:halpha_sdss}
\end{figure*}

Figure \ref{fig:halpha_sdss}a shows total H$\alpha$ emission contours overlaid on a SDSS i-band image.  The two galaxies and the area near the extragalactic HII region can clearly be distinguished as regions with the brightest H$\alpha$ emission, with extended emission spread between the galaxies.  For Taffy-S, the brightest H$\alpha$ lies on either side of the nucleus with fainter emission from the direction of the nucleus. It is noticeable that there is no obvious ionized gas associated with the faint southern part of the outer ring in Taffy-S, although there may be some associated with the northern part of the ring between the galaxies. Taffy-N has a very different distribution of ionized gas,  with a strong concentration in the inner disk, and fainter emission extending along the north-west major axis where it appears to join with bridge material. 

\begin{figure}
\centering
\plotone{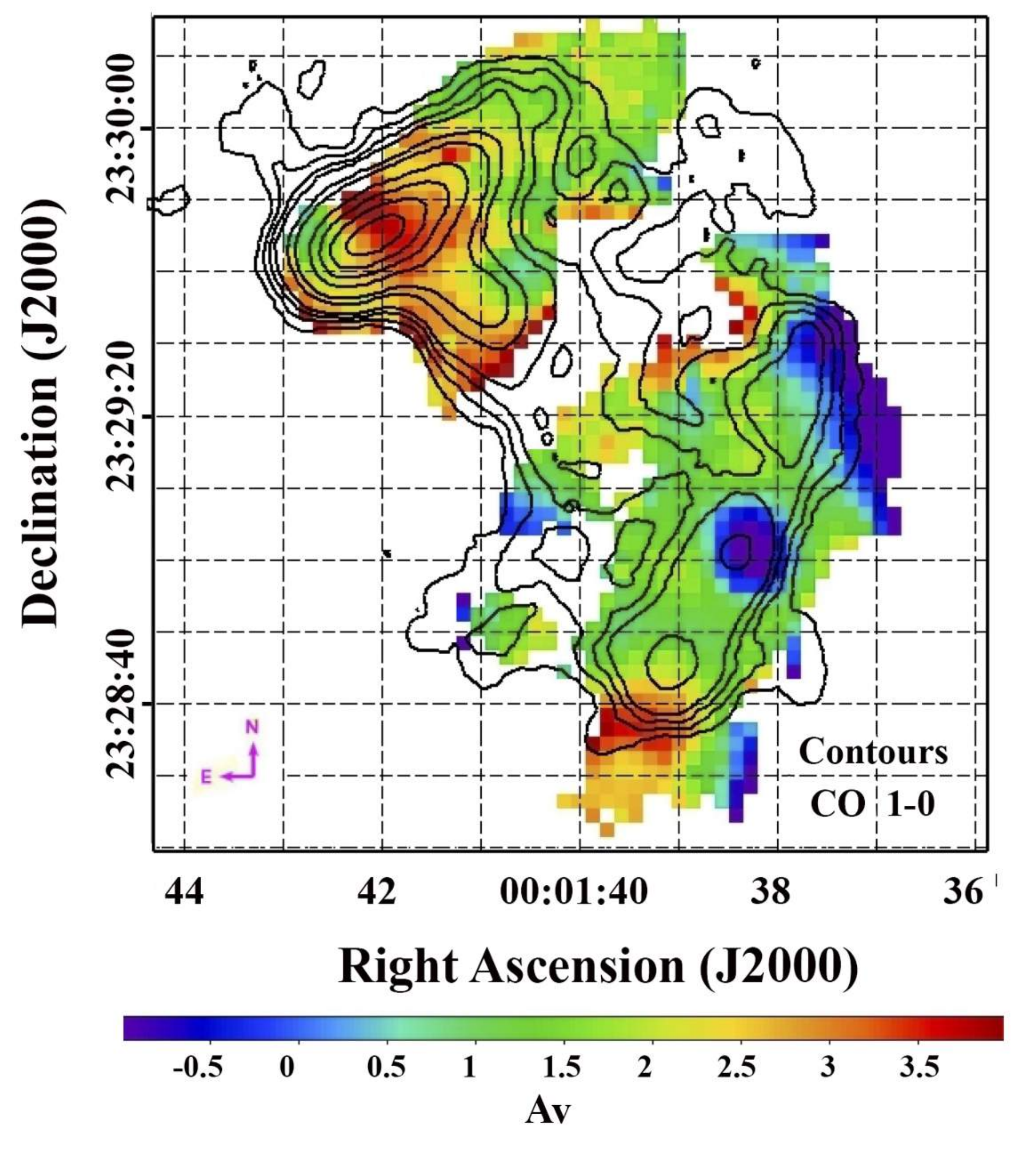}
\caption{Map of visual extinction, A$_\mathrm{V}$, with superimposed CO (1-0) (black) contours from \citet{Gao2003}. The A$_\mathrm{V}$ map was created as per the calculation described in \S\ref{sec:dust}. The contour levels are 35, 40, 45, 50, 65, 75, 100, 120, 140, and 160, in units of [Jy \kms beam$^{-1}$].}
\label{fig:av_map}
\end{figure}

Figure \ref{fig:halpha_sdss}b shows a map of the equivalent width of the H$\beta$ absorption across the system. In contrast to the lack of emission lines in the southern ring of Taffy-S, strong stellar absorption is seen there, along the extreme north-westerly edge of the galaxy, and in an extended region north-east of the faint stellar ring. In Taffy-N the strongest absorption is seen at the south-eastern end of the major axis in the region outside the main body of H$\alpha$ emission, although it extends at lower equivalent width as far as the center of the galaxy. We will discuss the implications of this H$\beta$ absorption in \S\ref{sec:hbeta_ew_results}.

\subsection{Dust extinction from the Balmer decrement}
\label{sec:dust}
We estimate the extinction caused by dust by examining the line ratio of the Balmer lines, H$\alpha$ and H$\beta$, referred to as the Balmer decrement, and assume Case B recombination. The color excess $\mathrm{E(B-V)}$ is given by,
\begin{equation}
\label{eq:color_excess}
\mathrm{E(B-V) = 1.97\, log_{10}\left(\frac{(H\alpha/H\beta)_{obs}}{2.86}\right)}.
\end{equation}

The extinction at wavelength $\lambda$ is related to the color excess by,
\begin{equation}
\label{eq:extinction}
\mathrm{A_{\lambda} = k(\lambda)\, E(B-V)}.
\end{equation}

We assume a reddening curve $\mathrm{k(\lambda)}$ of the form given by \citet{Calzetti2000}. Adopting the same method for all the observed lines allows us to correct, spaxel by spaxel, the observed line fluxes for dust extinction to arrive at intrinsic line fluxes. 

\begin{deluxetable*}{ c | c c c c c c c c }
\tablecaption{Visual dust extinction corrected emission line fluxes divided by region  [10$^{-16}$ W/m$^2$]. \label{tab:flux_table}}
\tablehead{
Region\tablenotemark{a} & A$_\mathrm{V}$ & H$\beta$ & $\mathrm{[OIII]\lambda5007}$ & $\mathrm{[OI]\lambda6300}$ & H$\alpha$ & $\mathrm{[NII]\lambda6583}$ & $\mathrm{[SII]\lambda6716}$ & $\mathrm{[SII]\lambda6731}$
}
\startdata
Taffy-N & 2.29 & 3.2  $\pm$ 0.23 & 1.98 $\pm$ 0.2  & 0.59 $\pm$ 0.16 & 9.12 $\pm$ 0.16 & 4.34 $\pm$ 0.14 & 2.24 $\pm$ 0.14 & 1.35 $\pm$ 0.16  \\
Taffy-S & 0.94 & 1.67 $\pm$ 0.13 & 0.67 $\pm$ 0.11 & 0.39 $\pm$ 0.11 & 4.78 $\pm$ 0.11 & 2.70 $\pm$ 0.1  & 1.38 $\pm$ 0.1  & 1.18 $\pm$ 0.12 \\
Bridge  & 1.89 & 1.42 $\pm$ 0.19 & 1.48 $\pm$ 0.16 & 0.55 $\pm$ 0.14 & 4.05 $\pm$ 0.13 & 1.89 $\pm$ 0.11 & 1.25 $\pm$ 0.11 & 0.81 $\pm$ 0.13 \\
\enddata
\tablenotetext{a}{Fluxes quoted here are from the combined Taffy-N and bridge regions i.e., including the western parts defined in Figure \ref{fig:Fig3}.}
\end{deluxetable*}

A map of the dust extinction at visual wavelengths, $\mathrm{A_V}$, derived from the Balmer decrement is shown in Figure \ref{fig:av_map} and superimposed on the  map are CO (1-0) contours from \citet{Gao2003} using data from the Berkeley-Illinois-Maryland-Association (BIMA) interferometer. It can be seen that the dust extinction map follows reasonably well the CO (1-0) surface density map, with a peak at the center of Taffy-N (as expected since the H$_2$ surface density is very high there). A high value of extinction is also seen at the southern tip of Taffy-S. This extends beyond the CO column density contours, but we note that the VLA maps of HI in this region by \citet{Condon1993} show a peak column density there of 1.8 $\times$ 10$^{21}$ cm$^{-2}$, which would imply an $\mathrm{A_V}$ of $\sim$~1 ~mag integrated over a 18 x 18 arcsec$^2$ beam \citep{Guver2009}. 

Interestingly, we observe a low value of $\mathrm{A_V}$ at the center of the Taffy-S, and along the northern edge of the spiral arm of that galaxy. The latter result is consistent with a low H$_2$ column there, although the low $\mathrm{A_V}$ in the nucleus may imply different conditions in the gas excitation (deviations from the assumed Case B recombination--perhaps due to a low-luminosity AGN) or a significantly reduced dust to gas ratio there.

Given that dust opacity measurements depend strongly on galaxy inclination \citep[see e.g.,][]{Driver2007, Unterborn2008}, we caution that our dust extinction estimates for both galaxies should be treated as lower limits and that actual A$_{\rm V}$ values could be much higher. For example, \citet{Gao2003} estimate A$_{\rm V}$ values could be higher than 10 mag for both galaxies. Our low A$_{\rm V}$ values for the Taffy galaxies could be because measuring the dust extinction from the H$\alpha$ and H$\beta$ lines involved in the Balmer decrement effectively only probes the effects of dust superficially \citep[essentially a ``skin'' effect; e.g.,][]{Calzetti2001}. It also assumes a simplistic dust geometry -- a screen of dust between the observer and the Balmer line emitting regions. Such an assumption might not be true for the complicated kinematics within the post collision Taffy system.

\section{The Kinematics of the Taffy System}
\label{sec:kinematics}
As Figure \ref{fig:Fig3} has shown, the spectra are quite complex in the system, and so we present the kinematic results in two ways. Firstly, we present the channel maps of one of the lines (in this case H$\alpha$) to provide a large-scale view of the gas distribution as a function of radial velocity channel. Secondly, we explore the spatial distribution of the gas associated with different kinematic components, especially those associated with multiple lines.

\subsection{H$\alpha$ Channel Maps}
\label{sec:vel_map}
\begin{figure*}
\centering
\includegraphics[width=\textwidth]{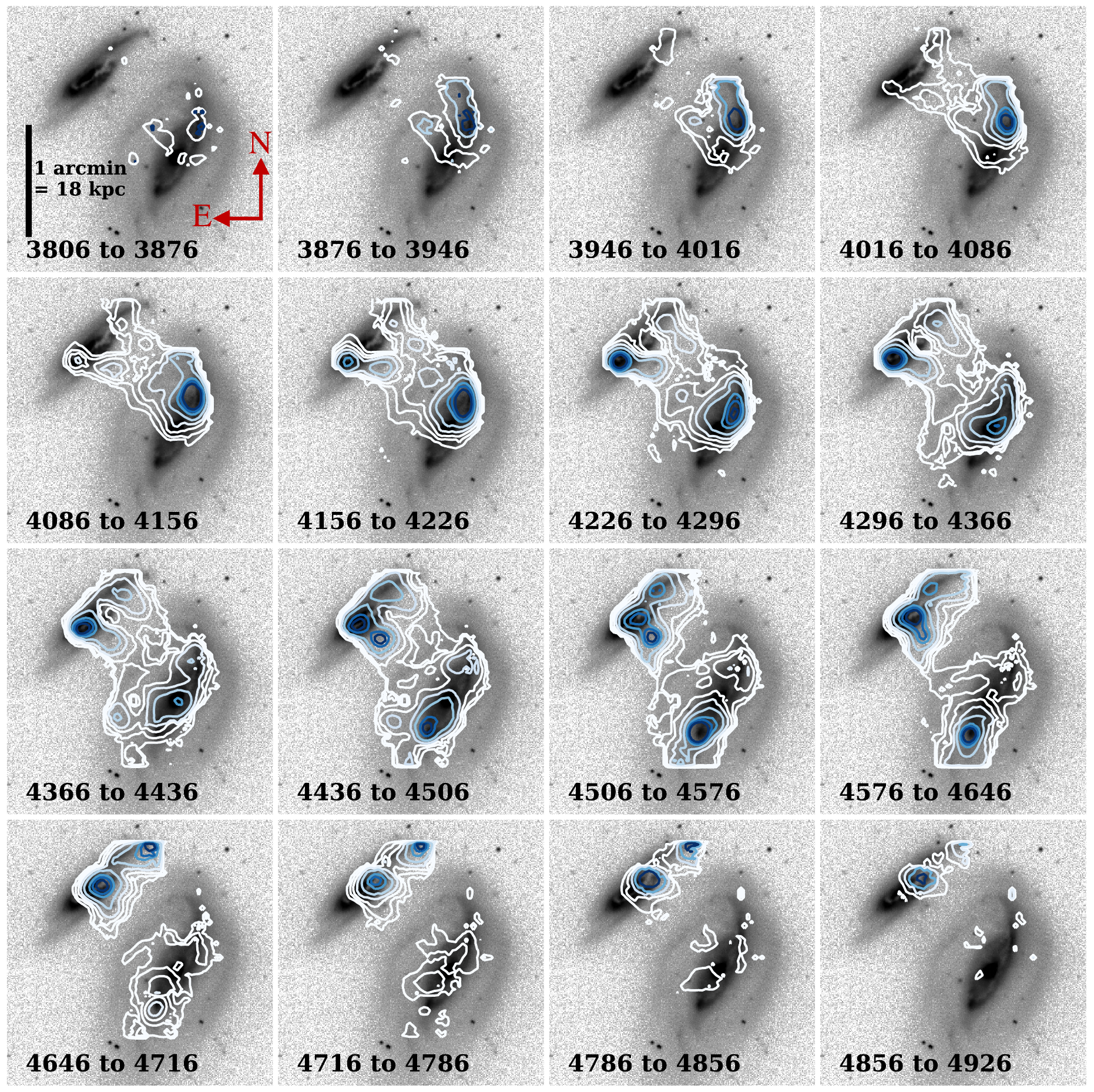}
\caption{Velocity channel map showing the H$\alpha$ emission over -540 to +510 $\mathrm{km\, s^{-1}}$ with respect to the (optically defined) systemic recessional velocity of $\sim4350\ \mathrm{km\, s^{-1}}$. The contours have been overlaid on a SDSS i-band image of the Taffy galaxies. The color of the contours, going from white to blue, indicates the intensity of the H$\alpha$ emission going from low to high intensity. The contour levels are 1, 4, 10, 20, 40, 80, 105, and 135 in units of ${\rm erg\, s^{-1}\, cm^{-2} / (70\, km\, s^{-1})}$. Note that every contour level is not visible in each panel because the H$\alpha$ intensities change for each panel. The velocity range shown for each panel (heliocentric velocity) is $70\ \mathrm{km\, s^{-1}}$. The LZIFU emission line cube was stitched as described in \S\ref{sec:methods_vel_comp}. The scale bar in the top left-hand panel is 1 arcmin (18 kpc for D = 62 Mpc) in length.}
\label{fig:vel_channel_map}
\end{figure*}

Figure \ref{fig:vel_channel_map} shows the H$\alpha$ emission channel maps integrated over channels of width  70~\kms. It is well known that the two galaxies are counter-rotating \citep[e.g.\ ][]{Vollmer2012}, and the brightest ionized gas in the two systems reflects this. Ionized gas in Taffy-S is seen in the lowest velocity channel (3806-3876~\kms)  in the north-west on the inside edge of the faint stellar ring, and progresses in a south-easterly direction with increasing velocity eventually showing a major component of emission in the south-east disk which fades away around 4780~\kms. Taffy-S also has some peculiar kinematics. For example, in the NW part of the disk, faint gas emission is seen over a wide range of velocities along the northern major axis even at the highest velocities. This would not be expected for gas in normal rotation.  Taffy-N is even more peculiar. The main centroid of emission from the south-east disk appears at around 4016-4086~\kms and progresses steadily towards the north-west, showing the counter-rotation. In addition, there is a peculiar region of emission which appears at even lower velocities on the north-west extreme tip of Taffy-N, and cannot be part of the normal rotation of the galaxy. Indeed, that structure appears to be part of the bridge, since as velocities increase it becomes more extended and eventually connects to the north-western region of Taffy-S. In addition to this bridge feature, a second bridge component  starts to appear between the galaxies at velocities of 4000~\kms, and at higher radial velocities it becomes quite strong in the region of the extragalactic HII region. The emission bridges the two galaxies where it joins with emission that potentially is associated with the faint stellar ring in the north-eastern part of Taffy-S. The connection between the galaxies disappears at velocities in excess of 4650 \kms.

The faint stellar ring in the northern half of Taffy-S exhibits some peculiar emission. Features that can be associated with this ring can be seen most clearly appearing from velocities around 4000~\kms. Moving to higher velocities shows several clumps that appear to follow the ring from NW to SE. These clumps also appear to be surrounded by emission that blends with the emission from the bridge. These features hint that this ring was strongly influenced by the collision and shows a discernible transition from material that is clearly associated with Taffy-S to material clearly associated with the bridge. A similar argument could also apply to Taffy-N although such a transition is much harder to observe in Taffy-N which is highly inclined. It is clear that the velocity structure of the gas in both galaxies and the bridge is very complex, and so we will now explore the gas in terms of its spectral profiles--which allows us to more easily separate normal regular rotation in the galaxies from peculiar motions.

\subsection{Mapping Two-component Line Profiles in the System}
\label{sec:methods_vel_comp}

As we have shown previously,  there are regions in the full data cube where the emission-line spectra show more than one component. Similar, double line-profiles were noticed in both the HI \citep{Condon1993} spectra, and in spectra taken in the far-IR [\ion{C}{2}] and [\ion{O}{1}] lines with {\it Herschel} \citep{Peterson2018}. Some regions of the bridge also show two components in the CO 1-0 molecular gas observations of \citet{Gao2003}. Our observations (consistent with those of \citealt{Gao2003, Peterson2018}) show that the double-line profiles in the ionized gas are not just confined to the bridge, but are also seen projected against parts of the galaxy disks, especially Taffy-N.  

To explore the kinematics further, we performed line fitting in two distinct passes.
Firstly we ran LZIFU, forcing it to fit only one component across the whole system. This worked well in regions where the lines were single-valued, but produced poor results in regions where the lines were double-profiled. The output from LZIFU at this stage was a model data cube built from the model fits, as well as the best fitting continuum cube. Also included were additional data products, including integrated line maps for each of the lines fitted.

Secondly, we ran LZIFU again, but this time we required it to fit two components for everything. In this case, the fitting worked well for the case of two components, but we found that it performed poorly in places where the profiles were singular. In this mode, since the software was fitting two separate line profiles, the output included two sets of data products (model line cubes, integrated line and velocity field maps), one set for each component--a low and high-velocity component. 

We were able to interrogate the model results to determine, spaxel by  spaxel,  the  mean, standard deviation, and amplitudes for each of the single and two component fits.  This allowed us to divide the results  into three classes of kinematic behavior: 

\vspace{0.4cm}
\begin{enumerate}
\item{Spectra consistent with a single Gaussian component, V$_s$ }  
\item{Spectra consistent with two components, V$_1$ and V$_2$, at different velocities  (V$_1$ represents the lowest velocity component, and V$_2$ the highest).}
\item{Spectra consistent with two components with nearly the same mean velocity, but exhibiting both a narrow V$_{1n}$ and broad  V$_{2b}$ component.}
\end{enumerate}
\vspace{0.2cm}
To qualify as two components, the two Gaussian line centers, V$_1$ and V$_2$, were required to differ by at least 35 \kms, or one-half of the velocity resolution at H$\alpha$ wavelengths. In practice, the lines were generally further apart than this and clearly separated.  Similarly, for a second component to be considered broad, V$_{2b}$ must have a FWHM of at least 1.5$\times$ that of V$_{1n}$. In a few cases, the LZIFU modeling failed to fit two components to cases where a single line profile would have been more appropriate.  In these small number of cases, we inspected the profiles by eye to perform what we considered to be the most reasonable  classification. 

In general when there was doubt about the classification, we inspected the profiles by eye to confirm the classification.  Overall the separation between single and two-components seemed reasonable, although we realize that there may be some areas where the distinction is somewhat subjective. Less subjective methods of deciding whether to fit single or multiple components to optical IFU data have been explored by  \citet{Hampton2017} using Artificial Neural Networks. These methods, which currently rely on training sets using "expert" astronomer guidance, are encouraging for the future, especially since such methods can classify velocity profiles faster than a human, and with statistically similar outcomes. As IFU data becomes more common, and as the amount of data increases with time, such methods may eventually be needed to replace human classifications. In our paper, in most cases, the classification of a two-component versus single Gaussian component was relatively unambiguous. 

In order to explore the relationship between the regions of emission where a single component is most appropriate compared with a region with two components, we have created composite moment maps (intensity, mean velocity and velocity dispersion) by {\it combining} those positions consistent with a single profile V$_s$ with those consistent with one or other of the double profile cases. The reason we combined the single component data with the two velocity components separately, was to look for continuity between the single component gas and one or other of the double-lines. For example, if the single component data mapped smoothly into velocity field of one of the two double components, this might suggest they are really part of one single dynamical system, whereas a sudden discontinuity would suggest no such regularity. 

\begin{figure*}
\centering
\includegraphics[width=0.95\textwidth]{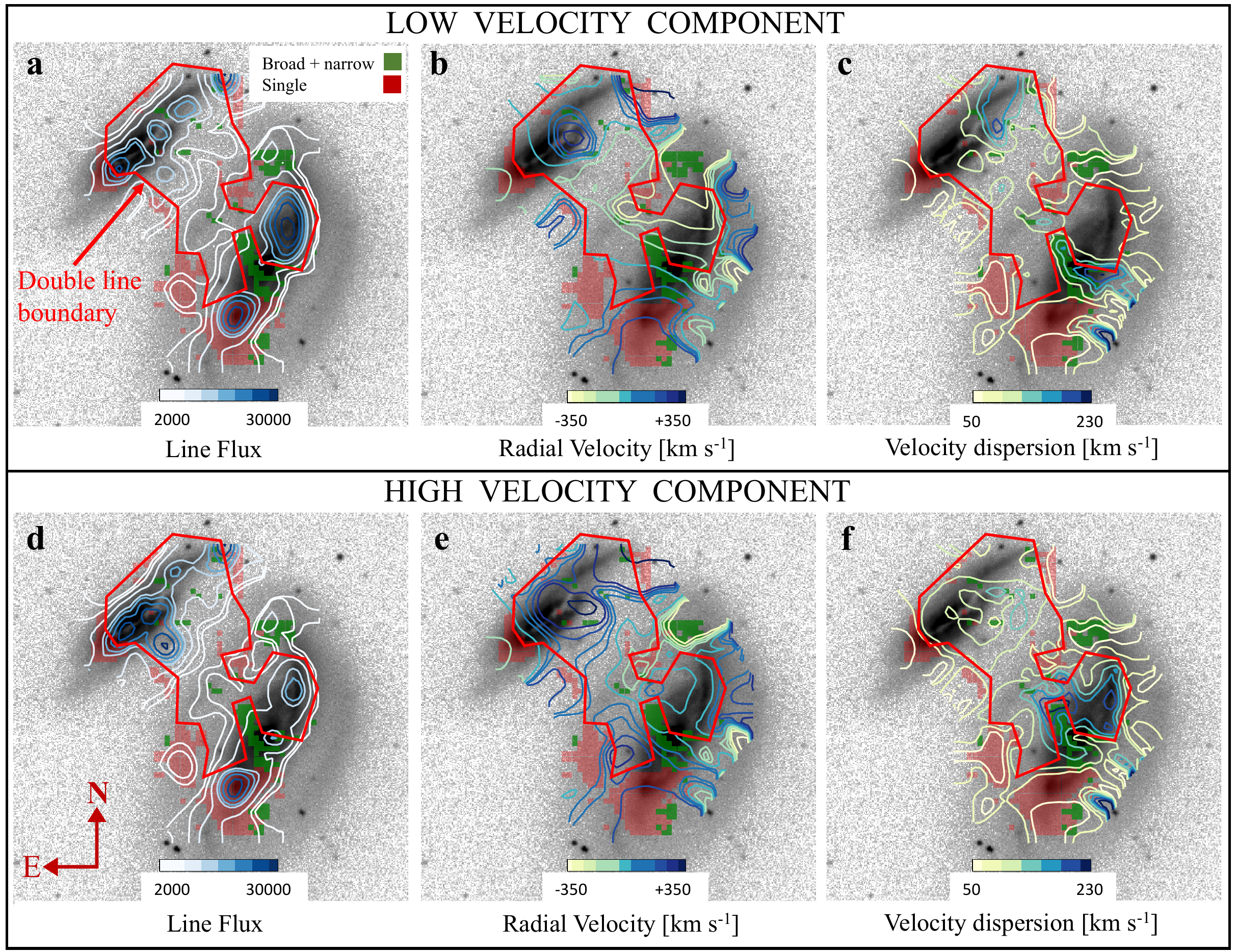}
\caption{Contour maps, overlaid on a SDSS i-band image of the Taffy system, of the moments of the velocity field. Zero velocity corresponds to an (optically defined) recession velocity of 4350 \kms. The top and bottom rows correspond to the low and high velocity component as shown. From left to right, the column panels show the integrated flux in H$\alpha$, radial velocity (with respect to systemic velocity), and velocity dispersion, respectively. The contour levels for the integrated line flux maps are: 2000, 3000, 6000, 12000, 15000, 20000, 25000, and 30000 in units of ${\rm erg\, s^{-1}\, cm^{-2}\, km\, s^{-1}}$. The contour levels for the velocity maps are: -350, -250, -200, -150, -100, 0, 100, 150, 200, 250, and 350 in units of ${\rm km\, s^{-1}}$. The contour levels for the velocity dispersion maps are: 50, 70, 90, 130, 160, 190, and 230 in units of ${\rm km\, s^{-1}}$. The red polygon demarcates the boundary where we see two line components in the profiles of the emission lines. The red spaxels outside the double line boundary indicate spaxels where we see only a single velocity component. The green spaxels indicate spaxels where we see two components but with significantly different widths (i.e., Class-3; we define the three line profile classes in \S\ref{sec:methods_vel_comp}).}
\label{fig:vel_moments}
\end{figure*}

These "merged" single and double profile moment maps are shown in Figure \ref{fig:vel_moments}.  Figures \ref{fig:vel_moments}~a,b and c represent moment maps created by combining spaxels containing components V$_s$ with V$_1$ and V$_{1n}$, whereas Figures \ref{fig:vel_moments}~d,e and f, represent the combination of spaxels containing V$_s$ with V$_2$ and V$_{2b}$.  To make it clear where the different kinds of profiles fall in the maps, we indicated in the Figure those regions enclosed within the red polygon that are consistent with Class-2 above (two components at different velocities). Those regions of the maps consistent with a single line, Class-1, are colored with a red background color.  Finally, those regions where a narrow and broad component were present, Class-3, are shown with a  green background color. 

Given that the kinematics are quite complex, we start by identifying those regions which may show regular galactic rotation. The simplest kinematics to understand are those of Taffy-S in the {\it low-velocity component} of Figure \ref{fig:vel_moments}b. Here, the increasing iso-velocity contours, going from yellow to dark blue progress regularly in the double-line region, merging smoothly with the V$_{1n}$ (green underlying color) and single-component V$_s$ (red underlying color) contours.  The velocity dispersion in the {\it low-velocity component} in Taffy-S is also low across most of its disk. Concentrating only on the {\it low-velocity component} for Taffy-S, it is clear that the velocities and dispersions shown in Figures \ref{fig:vel_moments}b and c look like a somewhat-warped, but regular rotating disk.  In contrast,  Taffy-S is much more peculiar in the {\it high-velocity component}  of Figures \ref{fig:vel_moments}e and f, where the galaxy shows only a small amount of obvious rotation, as well as exhibiting a high velocity dispersion in a large part of the disk. It also has a band of spectra classified as broad-line (Class-3 type; green underlying color) in the nuclear regions. 

We now turn our attention to Taffy-N which has more complex kinematics.  This galaxy is quite edge-on and may be expected to show regular rotation along its major axis. Evidence of rotation is seen in the {\it high velocity component}  of Taffy-N in the southern part of the disk centered on the dense dust lane and nucleus. Figure \ref{fig:vel_moments}e, shows a clear rotation signature where the velocities (starting with the pale blue contours in the south-east) increase along the major axis (dark-blue contours in the north-west of the inner disk). Although this apparent regular rotation in the  {\it high velocity component} is confined to the inner parts of the Taffy-N disk, the increasing trend in velocity shows a reversal towards the north-western extended disk. This might be interpreted as a turn-over in the rotation curve there.    

\begin{figure}
    \centering
    \plotone{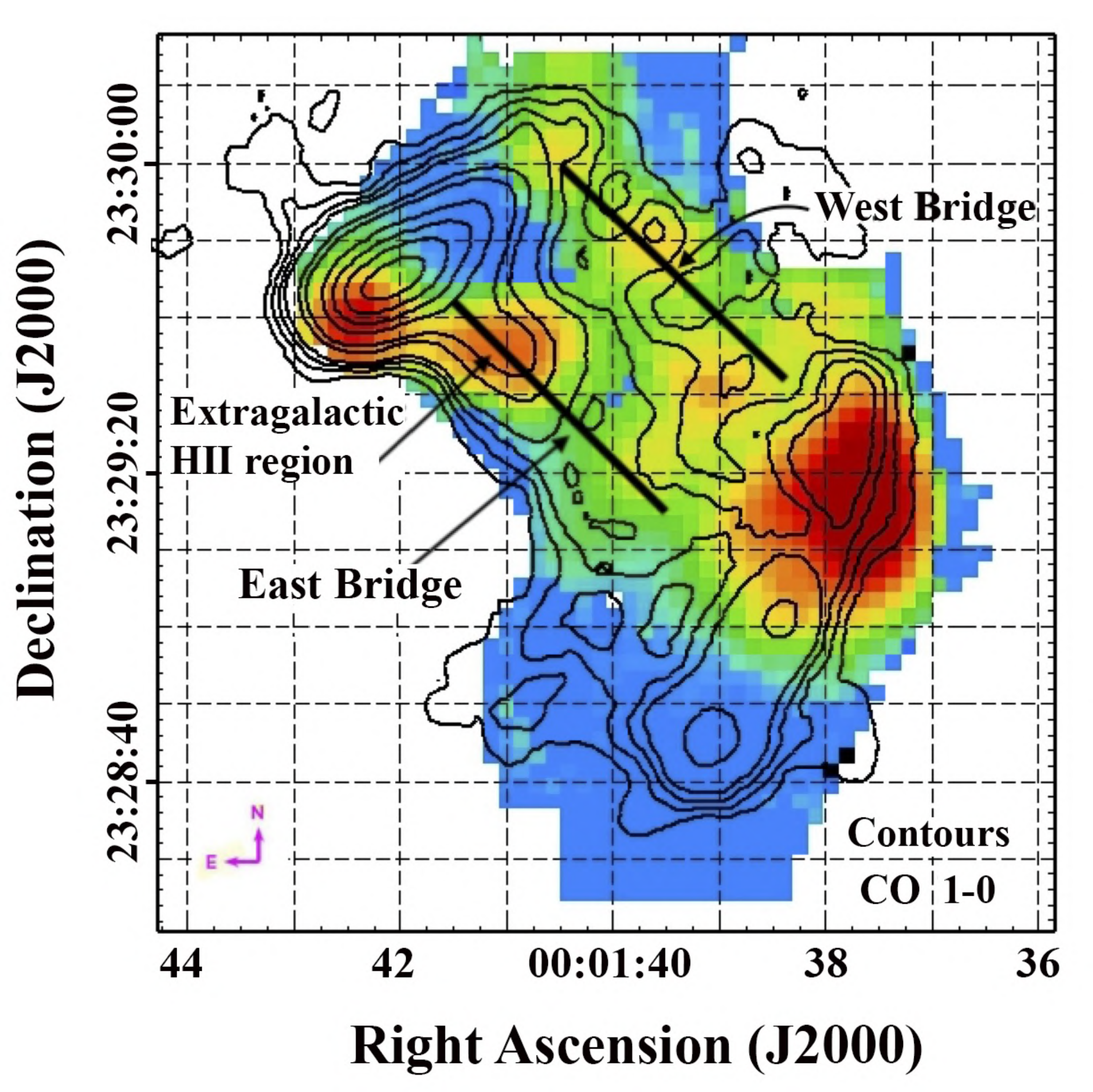}
    \caption{The emission between the galaxies can be decomposed into two kinematically different bridges of emission, seen represented in this single channel map of the H$\alpha$ for velocities 4156 to 4226 \kms  where parts of them both happen to appear at the same velocity. The two separate filaments are best defined by looking at the full range of channel maps shown in Figure \ref{fig:vel_channel_map}. The eastern bridge structure extends from the southern part of Taffy-N and extends down through the extragalactic HII region until eventually it merges with the south-eastern disk of Taffy-S. The western bridge extends from the north-west of Taffy-N into the bridge and eventually connects with the north-western tip of Taffy-S. The eastern bridge is more closely associated with the CO emission than the western bridge, although some clumpy regions are seen in CO even in the west (black contours are from \citet{Gao2003}, see text for more details).}
    \label{fig:summary_fig}
\end{figure}

Next we consider those parts of the velocity field that cannot be considered normal, and are most likely caused by the strongly collisional nature of the Taffy pair.  We have already pointed out in the discussion of the channel maps that the north-western part of Taffy-N has a peculiar low-velocity structure which is not part of the normal rotation. This can be seen in Figure \ref{fig:vel_moments}b ({\it low velocity component}) where much of the NW disk of Taffy-N shows very little rotation, and also shows high velocity dispersion (Figure \ref{fig:vel_moments}c). Here we see that the  gas extends as a finger towards the south-west, where it forms a western bridge with Taffy-S. A second eastern bridge structure, is seen associated with the extragalactic HII region, which is strongest in the {\it high-velocity component}. The two structures are graphically emphasized in Figure \ref{fig:summary_fig}.  Much of the gas between the two galaxies is seen in the eastern high-velocity bridge component, and shows a velocity gradient which extends between the two galaxies along direction of the radio-continuum bridge discovered by \citet{Condon1993}. Some regions of the high-velocity component bridge material have a high velocity dispersion--as was noted in the spectra in Figure \ref{fig:Fig3}.  We will show that much of the bridge gas in the {\it high velocity component} has the excitation properties consistent with shocked gas.

\begin{figure}
    \centering
    \plotone{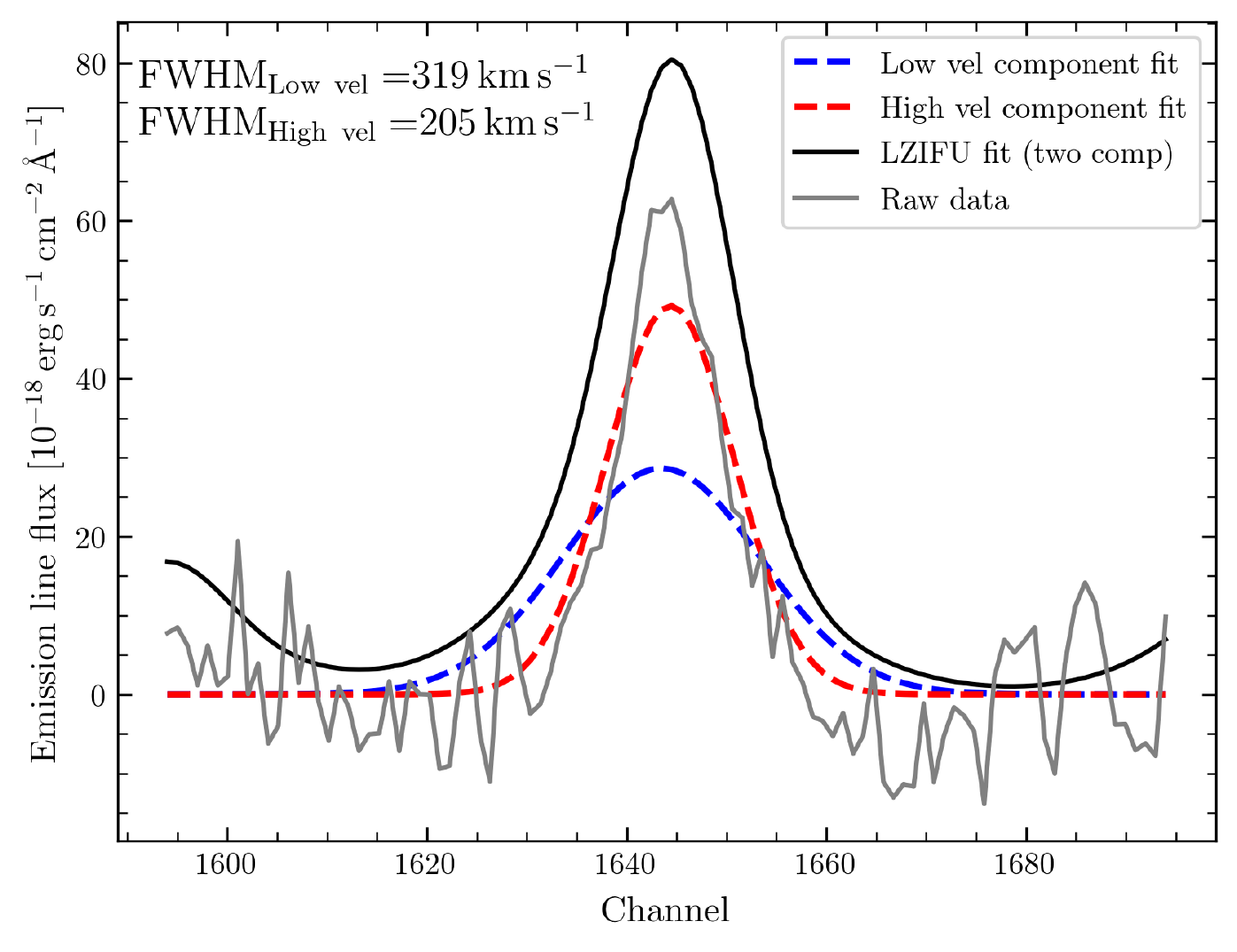}
    \caption{The H$\alpha$ nuclear spectrum of  Taffy-S (UGC~12914). The raw spectrum is shown (grey solid line), as well as the spectrum after correction by LZIFU for H$\alpha$ absorption (solid black line). We show the decomposition into two profiles, one with a broad lower-velocity component (blue dashed line) and a narrower slightly higher-velocity component (red dashed line)-see text. The spectrum emphasizes the importance of correcting for the continuum (driven mainly by the fit in the blue), since in this case the Balmer absorption masked a broader component in the emission line.}
    \label{fig:gaussfit}
\end{figure}

Several regions also show Class-3 spectra--which means that one component is broad. These regions (color green in all the panels of Figure  \ref{fig:vel_moments}) are confined to positions along the minor axis of Taffy-S, and to a small region at the north-west tip of the same galaxy. We show an example of these kind of spectra in Figure \ref{fig:gaussfit}. Here we show how LZIFU has fit two components to the H$\alpha$ profile from the nucleus of Taffy-S after correcting for weak Balmer absorption.  This appears as a region of high velocity dispersion in  Figure \ref{fig:vel_moments}c and f. The H$\alpha$ line profile shown is an average over a 4 x 4 spaxel region centered on the X-ray hot-spot in Taffy-S. The two components have a FWHM of 320 $\pm$ 70 \kms and 205 $\pm$ 70 \kms, with a small offset in velocity between the two. 

\section{Excitation of the Ionized Gas in the Taffy System}
\label{sec:bpt_results}
We next consider the possible excitation mechanisms for the ionized gas within the Taffy system by constructing emission line diagnostic diagrams based on LZIFU fitting of each spaxel in the data cube \citep[sometimes called BPT or VO diagrams;][]{Baldwin1981,Veilleux1987,Kewley2001}.  We construct emission line diagnostic diagrams using the [OIII]$\lambda$5007/H$\beta$, [NII]$\lambda$6583/H$\alpha$, [OI]$\lambda$6300/H$\alpha$, and [SII]$\lambda,\lambda$6716,6731/H$\alpha$ ratios. We have significant detections  of the [NII] and [SII] lines in most spaxels that fall on the galaxies and bridge, whereas the [OI]$\lambda$6300\AA\ line is detected in fewer spaxels because it is not as strong. We require every emission line used in the diagrams to be detected at the 3$\sigma$ level. 

In order to look for differences in excitation properties between the low and high velocity components where the lines are double, we show line diagnostic diagrams for each of the components separately (top panels a and b) in each of the Figures \ref{fig:nii_bpt}, \ref{fig:oi_bpt}, and \ref{fig:sii_bpt} for the [NII], [OI], and [SII] plots respectively. We also show the diagnostic diagrams for the total emission (sum of the two components plus those fitted by a single line) as a third panel (c) in each of the same Figures. These diagrams use classifications from \citet{Kewley2006}. In all three line diagnostic diagrams, we plot spaxels associated with different spatial regions of the Taffy-system. The symbols shown in the excitation diagrams are labelled in each Figure based on the regions defined in Figure \ref{fig:Fig3}.  For example, green points and crosses represent regions in Taffy-N, while blue points represent Taffy-S (the nuclear region is distinguished as blue diamonds).  The Taffy bridge is shown as red crosses (east bridge), and filled orange circles (west bridge). Because the west bridge contains fewer points than its eastern counterpart, we also present the integrated line ratio for the western bridge as a single larger open orange circle. 

The line diagnostic diagrams show that the low and high velocity components often behave differently, indicating a difference in their respective excitation mechanisms. For the two galaxies (blue and green symbols), all three sets of line diagnostic diagrams show spaxels mainly distributed within the HII, HII+AGN-composite or LINER part of the diagnostic diagram, with little hint of any pure AGN component. Much recent work has shown that excitation by shocks can resemble excitation by an AGN. \citet{Rich2011, Rich2014, Rich2015} showed that composite line ratios i.e., HII + AGN, can be due to HII + shocks.  This is particularly true for merging galaxies, for example \citet{Rich2014} showed that merging U/LIRGS can present ``composite'' optical spectra in the absence of any AGN contribution, with increasing contribution from shocks as the merger stage progresses from early to late-mergers. We will argue below that much of the LINER emission, with the possible exception of the nucleus of Taffy-S, is likely the result of fast shocks exciting the ionized gas in the bridge and in significant parts of the galaxy disks.

We will first concentrate on describing excitation in Taffy-N. We decided to split Taffy-N into two parts, one part covering the main disk of the galaxy and the other part covering the western extension which appears to connect with the western bridge. The emission from the main disk of Taffy-N (green points) is largely consistent with emission from HII-regions. This is true in all of the diagnostic diagrams in the low, high, and summed components. The western extension of Taffy-N shows a mix of HII and LINER emission. This is especially evident in Figure \ref{fig:oi_bpt} where the western extension of Taffy-N falls clearly in the LINER area in the low velocity component which we have previously noted from the channel maps may be associated with the western bridge.

Next we consider Taffy-S (blue points and diamonds). There appears to be a strong mix of HII-region and LINER/composite excitation for Taffy-S in all diagnostic diagrams. Despite the fact that the nucleus of Taffy-S does not show evidence for a powerful AGN in the gas excitation diagrams, this does not necessarily preclude a low-luminosity active nucleus being present-especially given the broader line-widths discussed previously. Taffy-S's nucleus (blue diamonds on Figures \ref{fig:nii_bpt}, \ref{fig:oi_bpt}, \ref{fig:sii_bpt}) does show some evidence of being a LINER, especially in the low-velocity regime. We find that most of the spaxels located on the nucleus of Taffy-S in the low-velocity (V$_1$) line diagnostic diagrams are in the LINER area in the [OI] and [SII] diagnostic diagrams, and close to the AGN line in the [NII] diagram for both the low and high velocity components. However, as previously noted, the velocity difference between the two components is small ($<20$ \kms), with the main difference being in the width of the lines, the V$_1$ component having a broader width than the V$_2$ component (see Figure \ref{fig:gaussfit}).   This is consistent with Chandra X-ray observations \citep{Appleton2015}, which showed the possible existence of a low-luminosity AGN based on the X-ray hardness ratio.

We divide the bridge into two parts (east and west) as discussed previously. The eastern bridge (red crosses) shows clear evidence of being HII-region excited in the low-velocity component. The situation is quite different for the {\it high-velocity V$_2$ component} (`b' panels) in Figure \ref{fig:nii_bpt}b, \ref{fig:oi_bpt}b.   Here, we observe sets of  east-bridge spaxels that deviate strongly from the HII area locus. For example, in the [OIII]$\lambda$5007/H$\beta$ ratio,  the east-bridge points are spread out along the [NII]/H$\alpha$ and [OI]/H$\alpha$ line, extending strongly into the LINER area. The western bridge (orange filled circles) show a mix of HII-region and composite/LINER behaviour in all diagrams. This is quite similar to what we find for the western extension of Taffy-N indicating that they might be excited by the same processes. Because there are so few points from the western bridge, we also plot an average of the entire western bridge as an orange open circle which is only shown in the `all' components diagram. This falls in the composite/LINER area for the [NII] and [OI] diagrams but in the HII area for the [SII] diagram.

\subsection{Evidence for shocked gas in the Taffy system}
\label{sec:shocks_discussion}
\begin{figure*}
\centering
\includegraphics[width=0.49\textwidth]{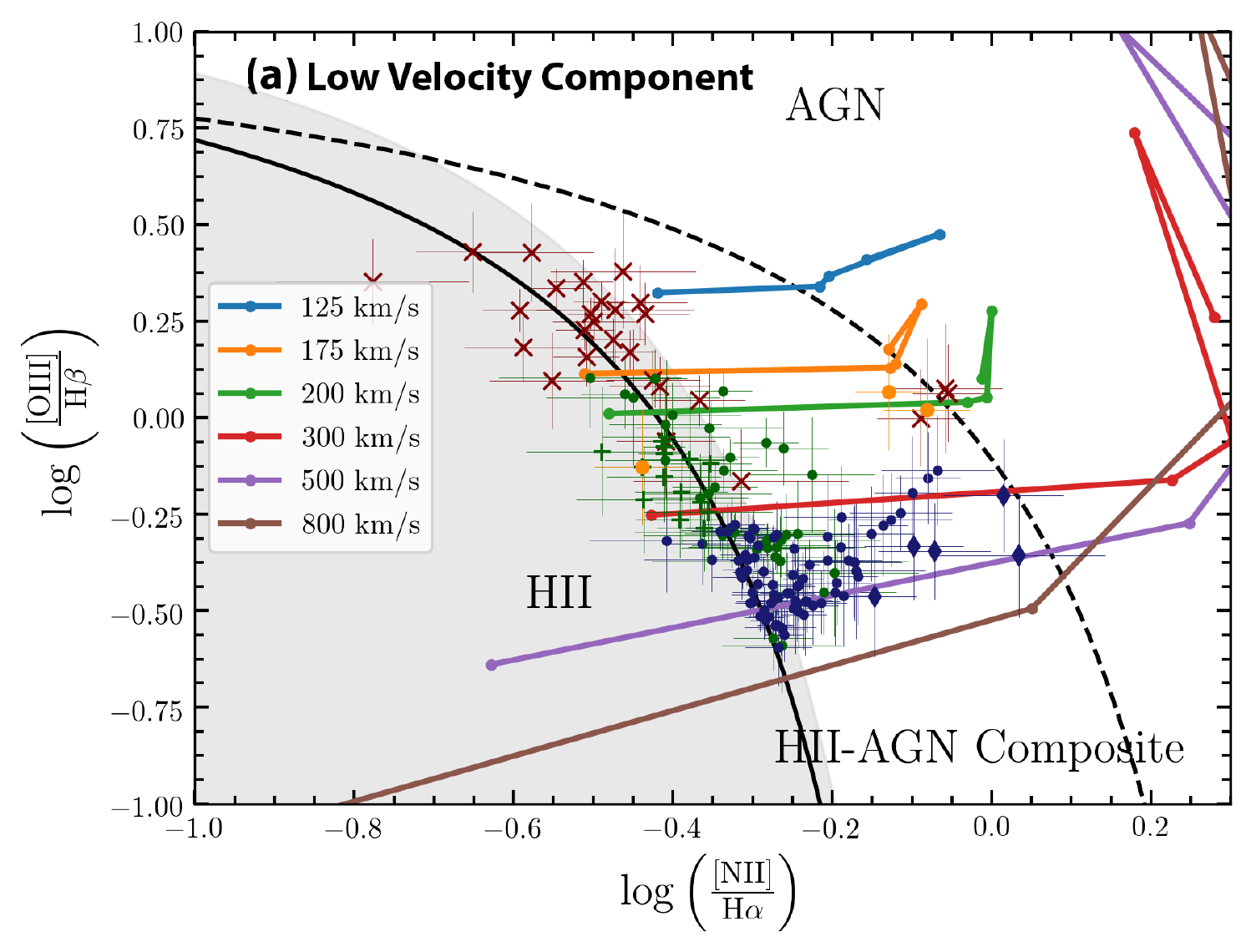}
\includegraphics[width=0.49\textwidth]{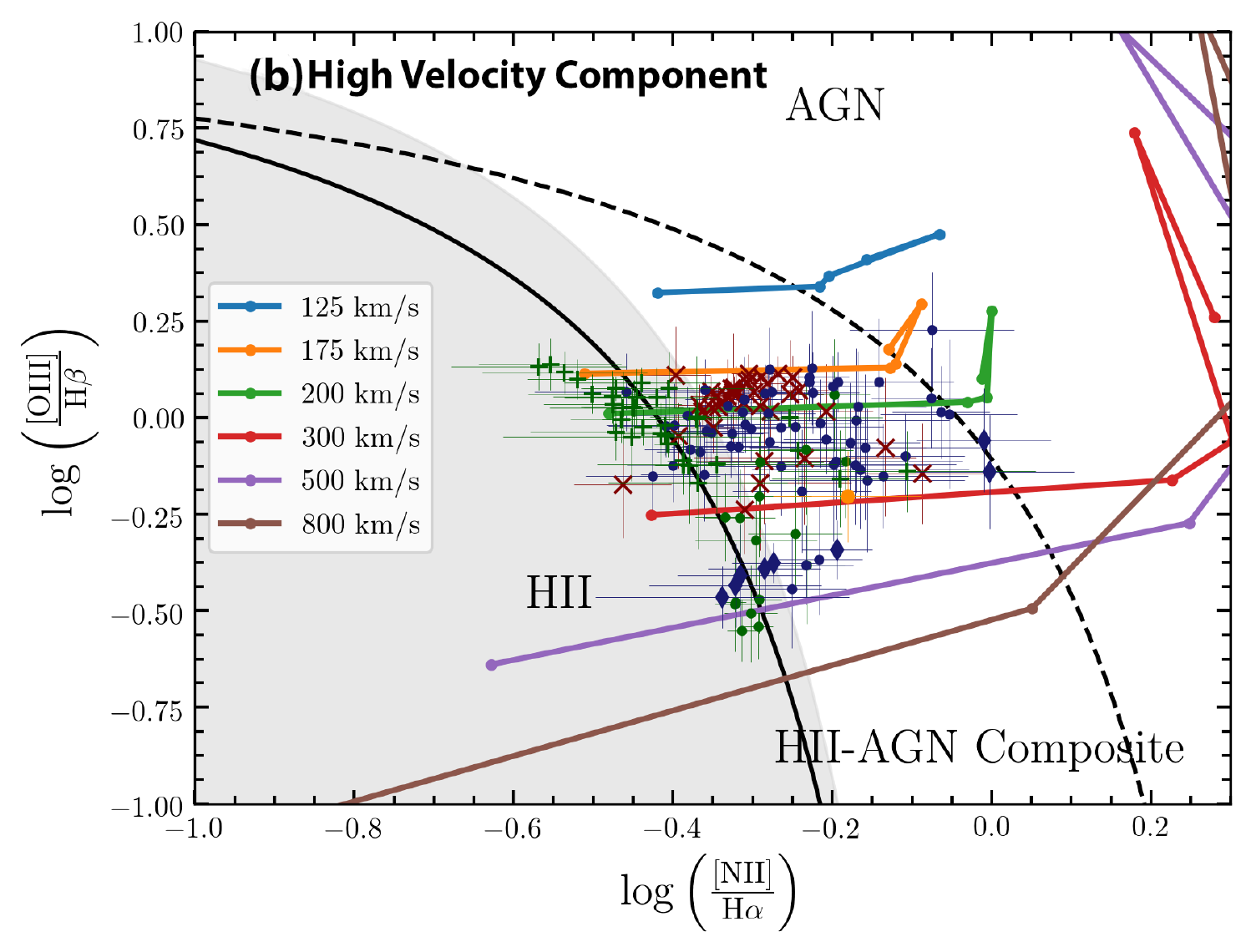}
\includegraphics[width=0.49\textwidth]{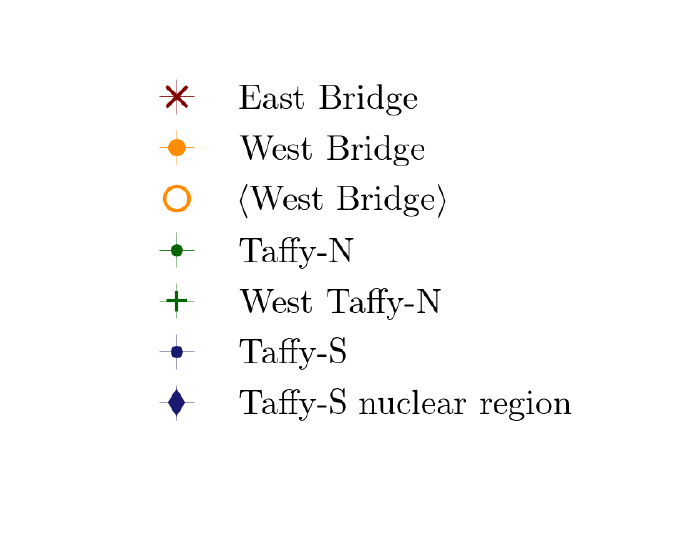}
\includegraphics[width=0.49\textwidth]{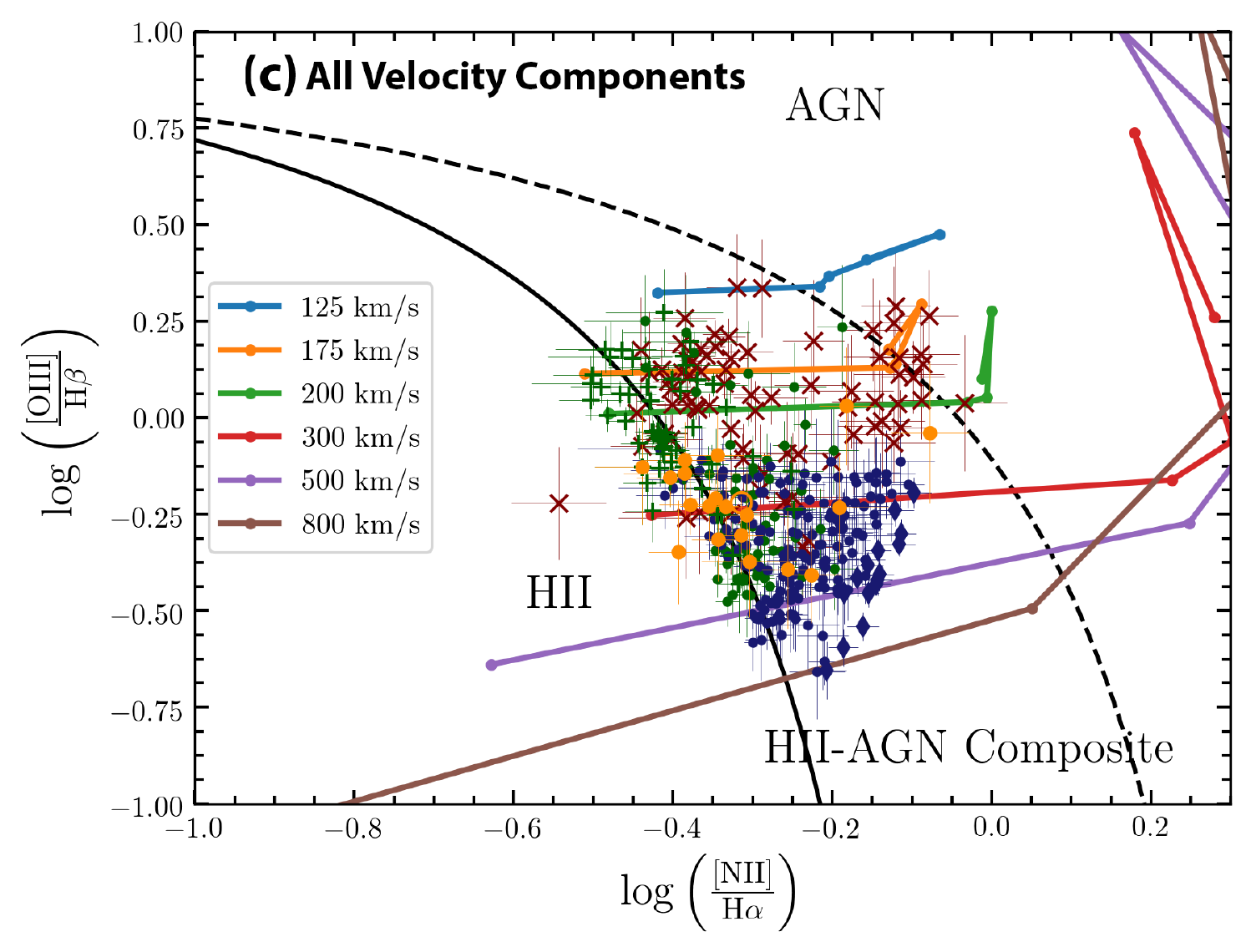}
\caption{[NII]$\lambda$6583\AA\ line diagnostic diagrams. (a): for the low velocity component, (b): for the high velocity component, and (c): total, i.e. sum of both velocity components and also including line ratios from spaxels which show a single component. Each point here represents the line ratios from a single spaxel. The red crosses, green points, and blue points correspond to the eastern bridge, Taffy-N, and Taffy-S, respectively. The blue diamonds are spaxels that fall within the nuclear region of Taffy-S. The green pluses are spaxels from the western part of Taffy-N. The orange circles are spaxels from the western part of the bridge. The unfilled orange circle is the average of line ratios from all the spaxels within the western bridge. These colors are consistent with those used to denote the corresponding regions in Figures \ref{fig:Fig3} and \ref{fig:summary_fig}, with the exception of the western part of Taffy-N which is shown as a magenta polygon in Figure \ref{fig:Fig3}. The panels also show line ratios from the MAPPINGS III shock models \citep{Allen2008} overlaid on the measured line ratios i.e. colored solid lines. The parameters assumed in the models are Z=Z$_\odot$ and B=5\,$\mu$G. Along each shock velocity line the points are marked by increasing number density. The classifications are from \citet{Kewley2006}. The shaded gray area around the solid black line classifying the HII-region excited gas marks the area we used to put a lower limit on the fraction of gas excited by star formation (see \S\ref{sec:frac_from_sf}). The width of the shaded area (above the HII classification line) is twice the size of the average error bar in each panel.}
\label{fig:nii_bpt}
\end{figure*}

\begin{figure*}
\centering
\includegraphics[width=0.49\textwidth]{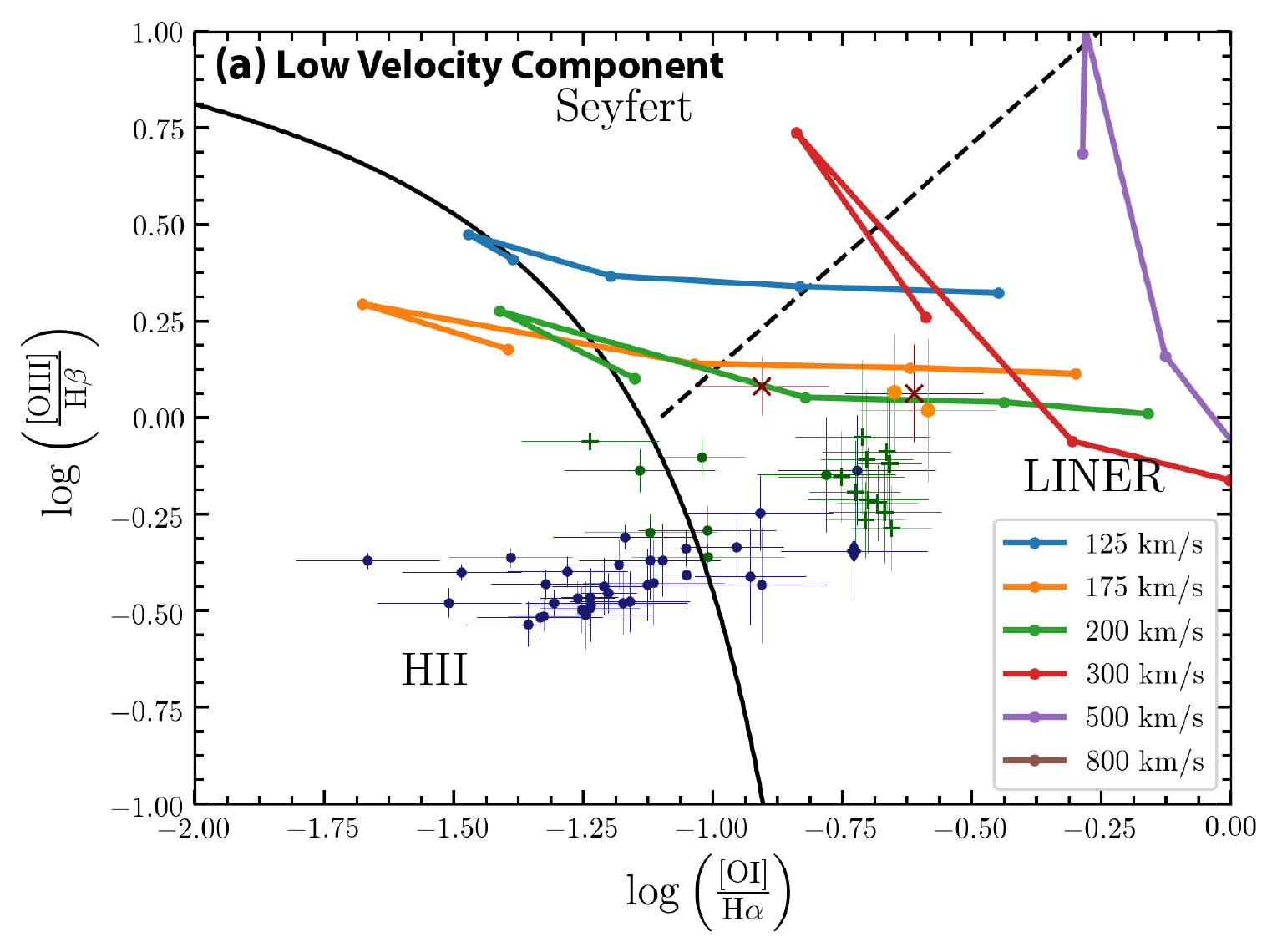}
\includegraphics[width=0.49\textwidth]{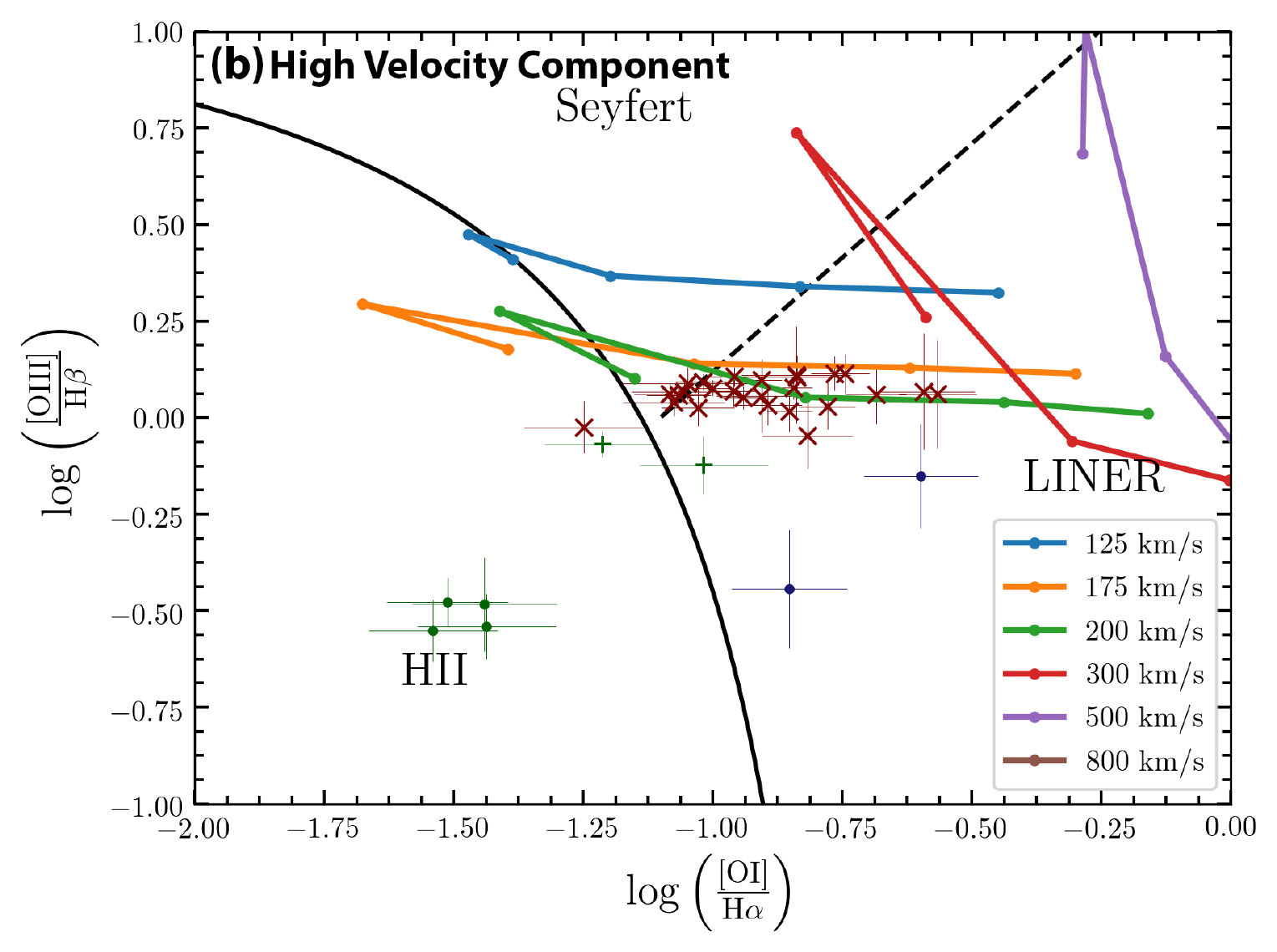}
\includegraphics[width=0.49\textwidth]{emissionline_diagnostic_legend.pdf}
\includegraphics[width=0.49\textwidth]{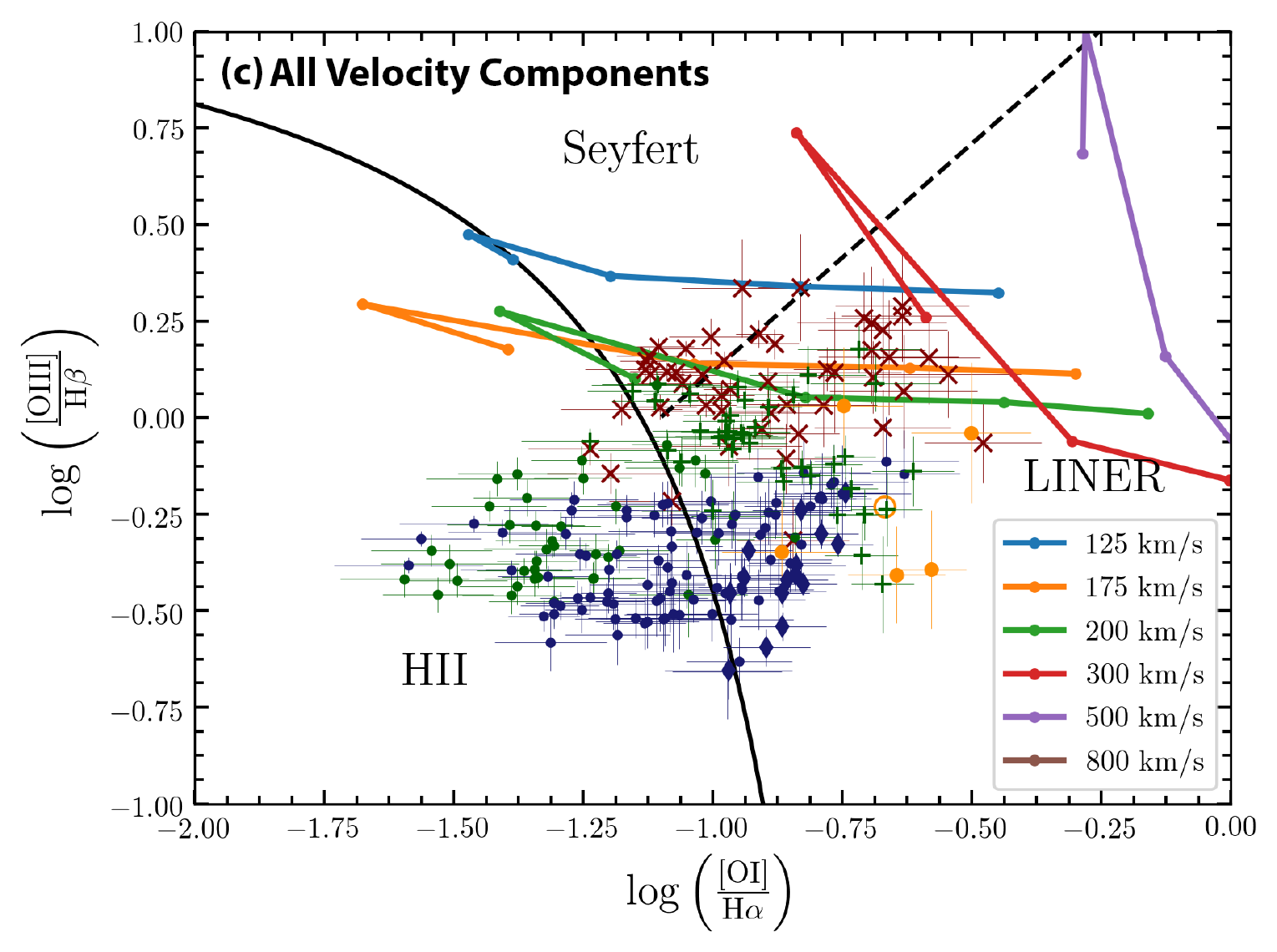}
\caption{Same as Figure \ref{fig:nii_bpt} but using the [OI]$\lambda$6300\AA\ line. This Figure contains fewer points than the [NII] and [SII] line diagnostic diagrams due to the [OI] line being much weaker than the [NII] and [SII] lines and therefore being undetected in many spaxels.}
\label{fig:oi_bpt}
\end{figure*}

\begin{figure*}
\centering
\includegraphics[width=0.49\textwidth]{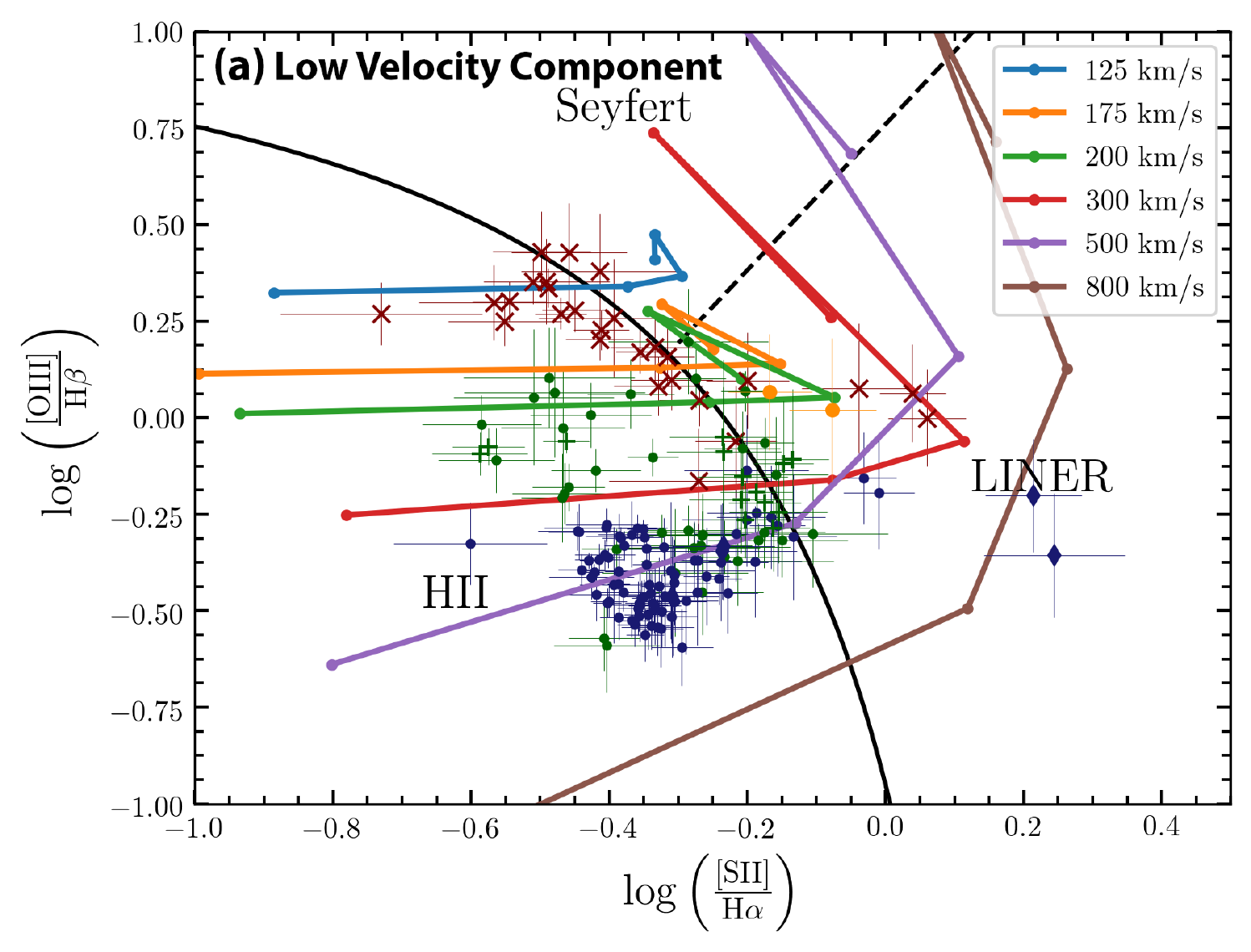}
\includegraphics[width=0.49\textwidth]{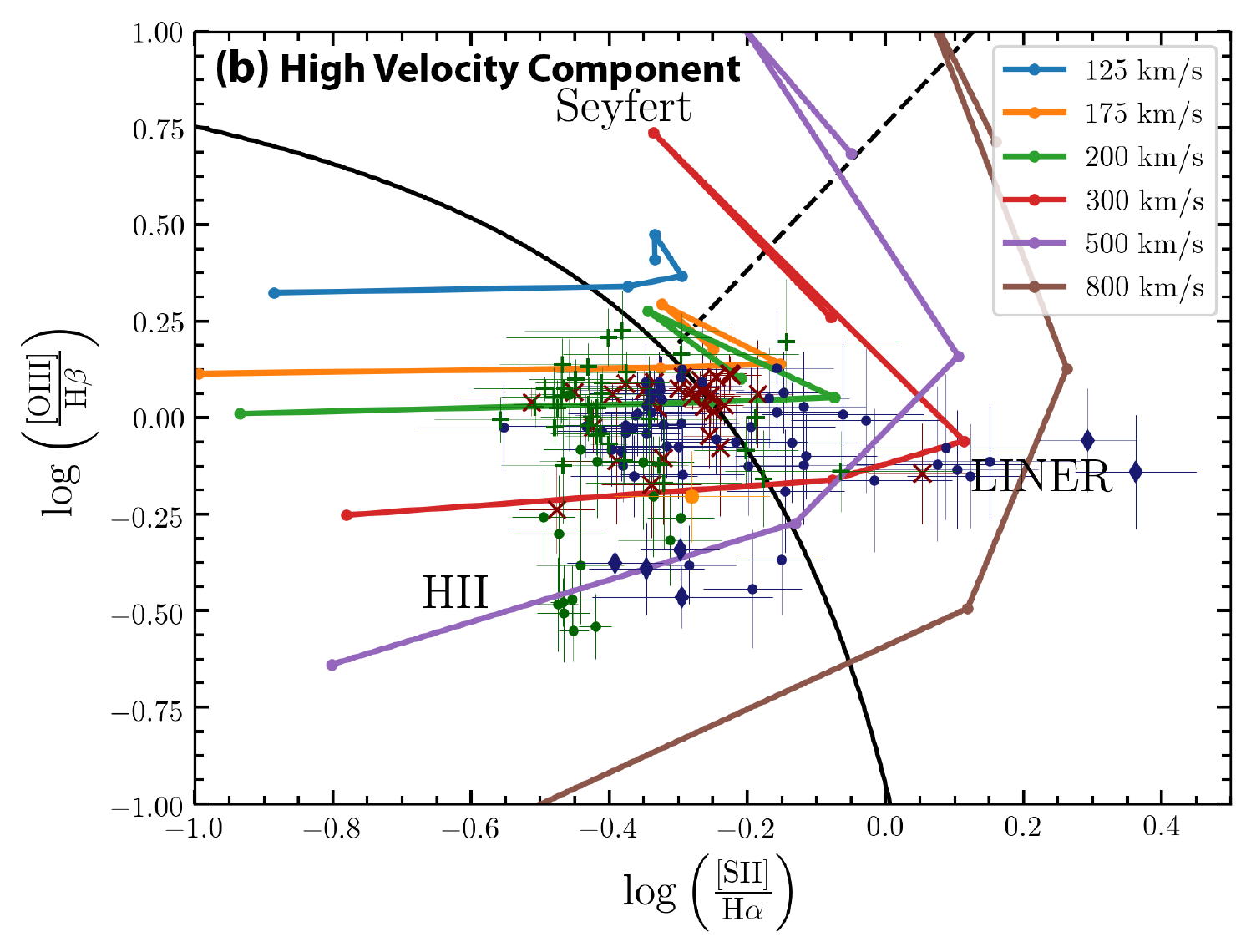}
\includegraphics[width=0.49\textwidth]{emissionline_diagnostic_legend.pdf}
\includegraphics[width=0.49\textwidth]{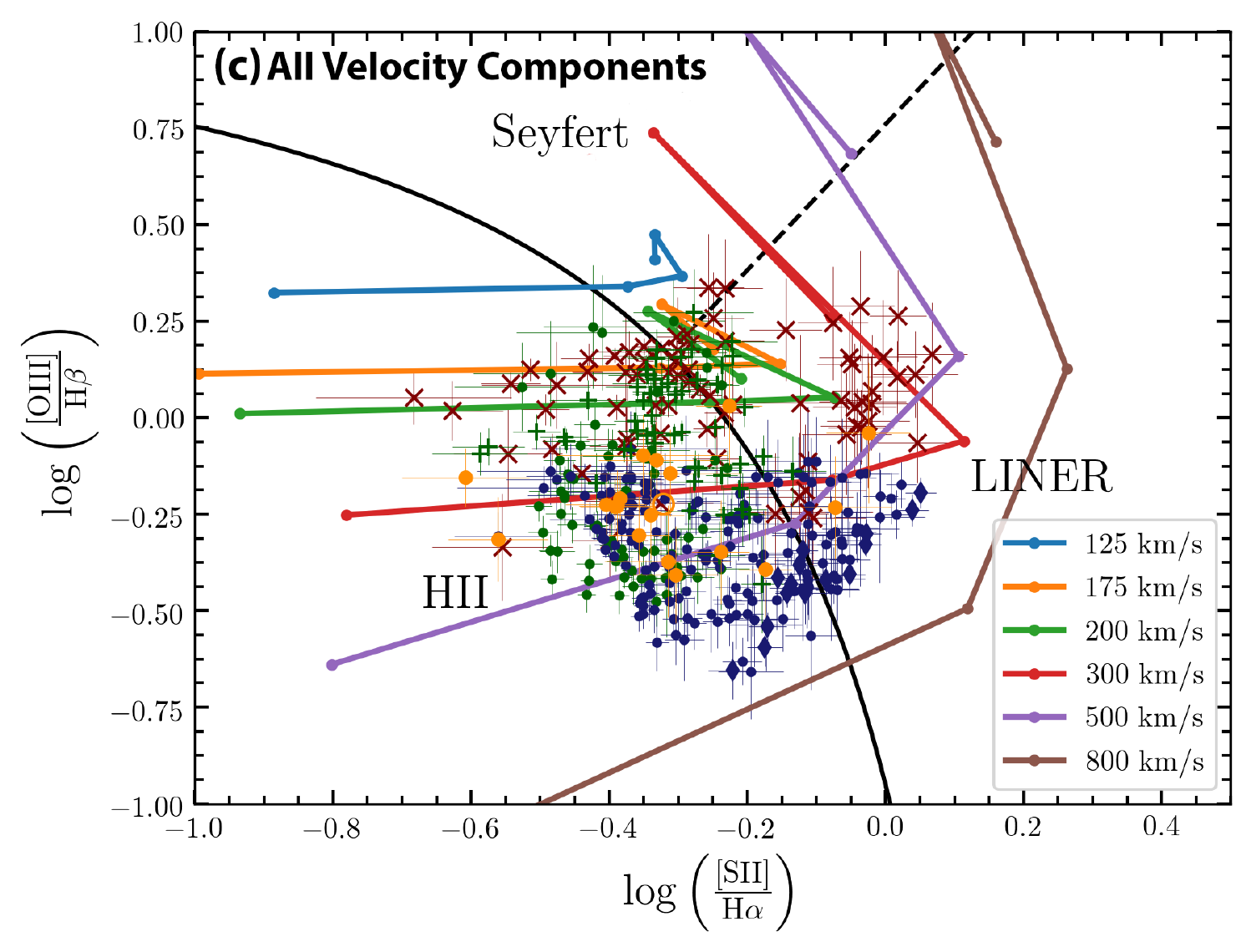}
\caption{Same as Figure \ref{fig:nii_bpt} but using the sum of the [SII]$\lambda$$\lambda$6716,6731 lines.}
\label{fig:sii_bpt}
\end{figure*}

We first explore the possibility that the gas is excited by shocks.  We over-plot predicted line ratios from shocks on Figures \ref{fig:nii_bpt}, \ref{fig:oi_bpt}, and \ref{fig:sii_bpt}, taken from the MAPPINGS III library of models \citep{Allen2008}. The model line ratios are plotted for different shock velocities as solid colored lines. Based on the models of \citet{Vollmer2012}, we assume that much of the gas in the bridge and throughout the galaxies is close to solar metallicity, since the gas has been stripped from the galaxies or has been excited {\it in situ}.  For the shock models we therefore assume solar metallicity. The other parameters of the models include the pre-shock gas densities in the range $\mathrm{0.1 < n\, [cm^{-3}] < 1000}$ (stepping by a factor of 10 each time), and an assumed constant magnetic field of B=5 $\mu$G (this is close to the equipartition magnetic field strength of 8 $\mu$G measured by \citealt{Condon1993} through radio continuum measurements). We have plotted only the line ratios for the shock itself while excluding the precursor component of the shock. For the moderate shock velocities that seem compatible with the Taffy excitation velocities, it may be reasonable to ignore the effect of a strong ionizing shock-precursor, as we shall discuss later.

For the east bridge points (red crosses), in the high-velocity component (`b' panels), it is clear that the east bridge points fall relatively neatly between the solid lines for shock velocities 175 km/s and 200 km/s (green and orange lines). For the west bridge points, in cases where the points deviate from the HII-region area they are consistent with the same shock velocity, e.g., the low-velocity components in all diagrams. {\it Thus the high-velocity east bridge component and the low-velocity west bridge component seem consistent with shock excitation for all the line diagnostic diagrams.} 

The situation is mixed for the disks of galaxies themselves. As Figures \ref{fig:nii_bpt}b, \ref{fig:oi_bpt}b and \ref{fig:sii_bpt}b show, there are some points within the disks of both galaxies which fall in the composite region of the diagnostic diagram. Some of these points would be consistent with a mixture of HII region and shocked gas excitation. The nucleus of Taffy-S is an ambiguous case, because it could be excited by shocks in a mild outflow, or may be gas excited by UV emission from a weak LLAGN. 

The spreading of the points along lines of constant [OIII]$\lambda$5007/H$\beta$ ratio in the high-velocity component in the bridge has a number of possible interpretations if we assume that shocks are involved. Firstly, the spread might imply that the shocks are occurring in an ensemble of gas clouds with different pre-shock densities. Such a picture is consistent with our previous observations of the Taffy bridge \citep{Peterson2012, Peterson2018} where we have observed gas in many different excited phases, from HI \citep{Condon1993} to  warm molecular gas from the {\it Spitzer} IRS; along with the detection of  [CI]  and [CII] emission \citep{Peterson2018}, and boosted values of [CII]/FIR and [CII]/PAH ratios. The existence of a highly multi-phase (and multi-density) medium is also very consistent with the detection of soft X-ray emission from the bridge. Thus it might be expected that shocks moving through such a multi-phase gas would encounter a range of pre-shock densities--which would spread the points along lines of constant shock velocity--as observed in Figures \ref{fig:nii_bpt} and \ref{fig:oi_bpt} especially.

An alternative explanation might be that some of the gas is of lower metallicity. As the models of \citet{Allen2008} show, reducing the metallicity of the shocked gas moves the points in the diagnostic diagrams to the left at roughly constant values of [OIII]$\lambda$5007/H$\beta$ ratio. However, if the collisional models of \citet{Vollmer2012} are correct, the gas in the bridge should have come from many different places within the original pre-collisional disks, and deviations of factors of 100 in metallicity in the bridge seem unlikely.  We conclude that it is much more likely that we are observing shocks within the bridge and parts of the galaxy disks which encounter clumps of material at different densities.  

\begin{figure*}
\centering
\includegraphics[width=0.49\textwidth]{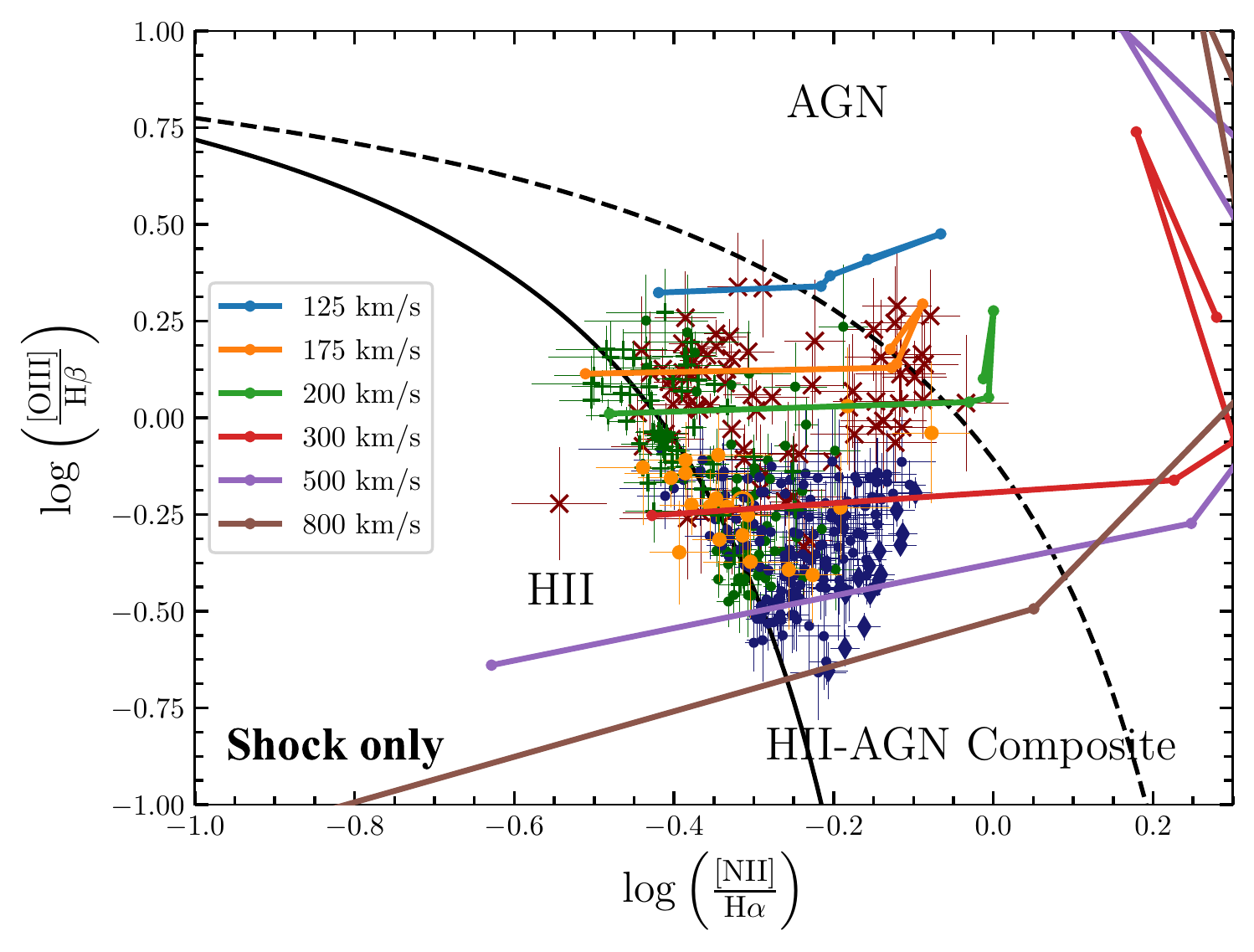}
\includegraphics[width=0.49\textwidth]{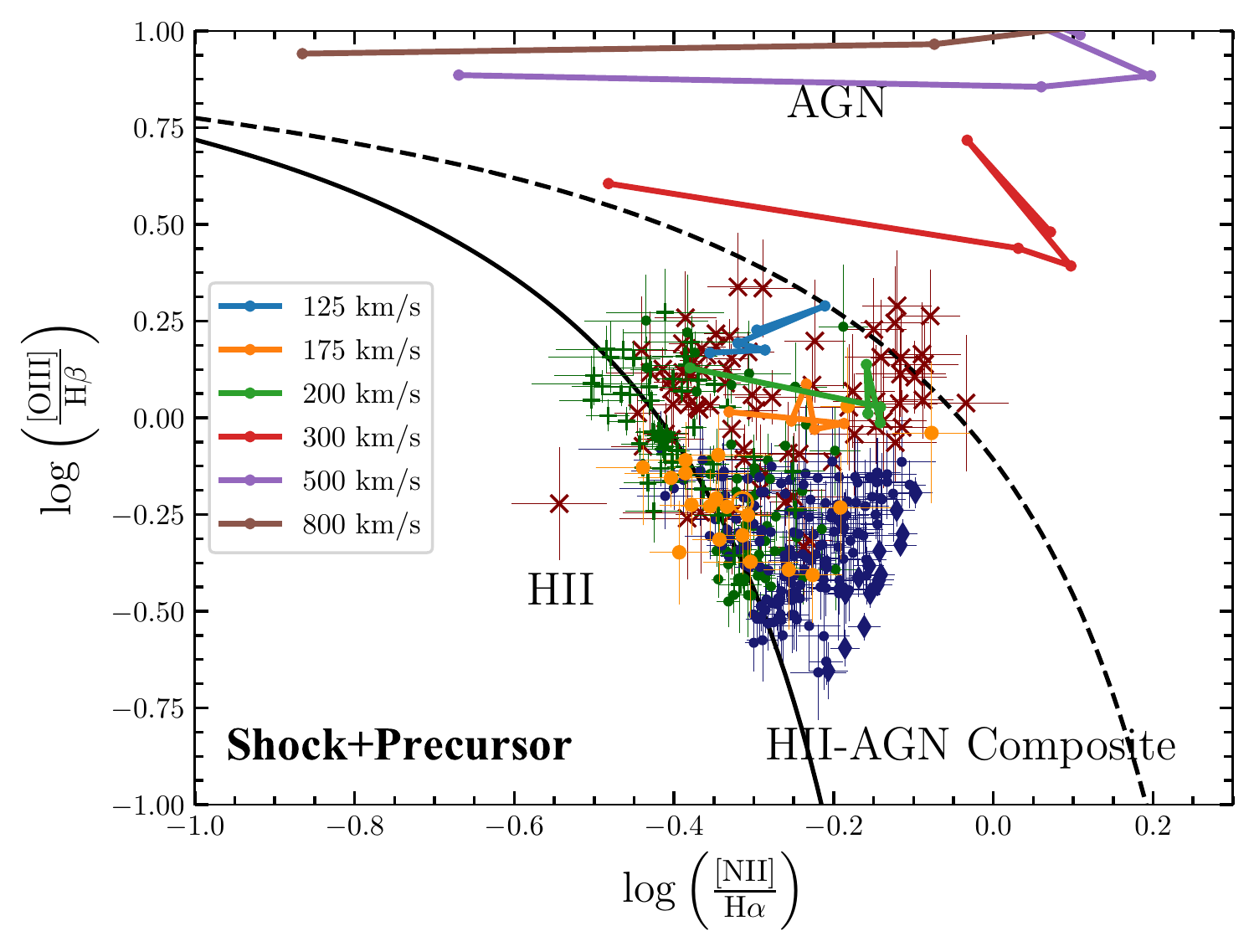}
\caption{The effect of including the only the shocks compared to including shocks with the shock precursor, from the models of \citet{Allen2008}, for the [NII] diagnostic diagram. The symbols (and their colors) and shock model parameters are the same as the line diagnostic diagrams of Figures \ref{fig:nii_bpt}, \ref{fig:oi_bpt}, and \ref{fig:sii_bpt}. The included data points are the same as those on the ``All Velocity Components'' diagram of Figure \ref{fig:nii_bpt}.}
\label{fig:Fig13}
\end{figure*}

What is the effect of ignoring the possible influence of a hot shock precursor in the models? This is an effect where, in high velocity shocks, the gas in the shock is so strongly heated that UV radiation from the shocked gas ionizes large amounts of pre-shocked gas upstream of the shock. We show in Figure \ref{fig:Fig13}, an example of the [NII] line diagnostic diagram, the effect of including the shock and the shock precursor \citep{Allen2008}. As can be seen by comparison with Figure \ref{fig:nii_bpt} the behaviour when we include the shock precursor with velocities $<300$ \kms\ is very similar to the case with no shock precursor, which fits our data well. At shock velocities $>300$ \kms\ the models including the shock precursor diverge significantly from these data. Similar behaviour is noted in the other diagnostic diagrams (not shown). This implies that the shock models between 100-300 \kms\ fit the bridge data well regardless of whether the precursor is included. We note, however, that in dense gas (e.g.\ molecular gas known to also be present in the bridge) the velocity at which a shock precursor may become important will be much lower. Therefore future modeling the molecular shocks may have to take precursor activity into account.

\subsection{Alternatives to Shock Excitation: Diffuse Lyman Continuum emission leaking from HII Regions?}

In a recent paper by \citet{Weilbacher2018}, it was noted that the Antennae galaxies (NGC 4038/39), like the Taffy system,  also exhibit significant diffuse ionized gas emission. Although several mechanisms were put forward to explain the emission, including the possibility of shocks, the authors favor an  interpretation that much of the diffuse gas is ionized by Lyman-continuum photons (hereafter Ly-C) "leaking" from large numbers of HII regions found primarily in the disks of both galaxies, as well as HII regions found within the "overlap region". Much (but not all) of the diffuse component was found close to massive star formation complexes in the system. Using multi-color HST imaging of the clusters, the authors were able to compare the luminosity of the H$\alpha$ emission associated closely with a cluster with theoretical models of the Ly-C flux from the clusters to determine whether the clusters were "leaking" Ly-C photons  into the surrounding gas.   It was found that many of the HII regions had non-zero escape fractions of Ly-C UV radiation, especially in the center of NGC 4038 and also in the "overlap" region. Thus, for the Antennae system, it was found that the excess diffuse emission within the system could be explained as gas excited by UV radiation escaping from the clusters. 

Could some, or all of the extended ionized gas emission we see in the Taffy come from similarly degraded UV light from HII regions in the disks of the Taffy galaxies that might diffuse outwards and ionize large parts of the Taffy bridge? We estimate (Table \ref{tab:flux_table})  that the amount of H$\alpha$ emission coming from the bridge, after correcting for extinction, is similar to that from Taffy-S and roughly 50$\%$ of the emission from Taffy-N. In this case the majority of the escaping photons would have to come from the galaxies themselves, since the star formation rate in the Taffy bridge is very low. Unfortunately, unlike the case of the Antennae, we do not have multi-color high resolution images of the individual star clusters, and this means that it is difficult to perform the same kind of test that was applied, for all individual HII regions, by \citet{Weilbacher2018}. As a result, we cannot completely rule out a significant contribution to the ionized medium in the Taffy coming from leaky HII regions. Nevertheless, many other lines of evidence already suggest that shocks must be present in the Taffy bridge, and so we prefer the shock explanation for the excitation of the high-velocity component, rather than leaky HII regions. Future observations will be needed to attempt to model the history of star formation in the clusters in the galaxies, which will allow us to estimate the fraction of UV emission which may escape into the surround gas. This is beyond the scope of the current paper.

\section{Ionized Gas Fractions, Star formation Rates and Mass in the ionized Component}
\label{sec:frac_from_sf}

Here we describe the method that we employed to estimate a lower limit to the fraction of ionized gas excited by star-formation (as opposed to being shock excited) using the [NII] line diagnostic diagram. We use the [NII] line diagnostic diagram since it contains the most number of points. We start by defining an effective HII-region excitation area. This is shown as a shaded gray region in the [NII] line diagnostic diagrams of Figure \ref{fig:nii_bpt}. We defined this area simply by ``padding'' the HII classification line \citep{Kewley2006} by twice the size of the average error in the y-direction (the y-error being the larger error). We sum up the H$\alpha$ flux in each spaxel that falls within this region. This flux is divided by the total H$\alpha$ flux to arrive at the lower limit for the fraction of ionized gas excited by star-formation. This process is repeated independently on each velocity component.

The fraction derived this way is a lower limit because the other spaxels, outside of the shaded area (i.e.\ in the HII+AGN area) will contain emission from gas excited by star-formation, and we cannot accurately disentangle the excitation from shocks and star-formation (also see text in \S\ref{sec:shocks_discussion}). The lower limits for the fraction of ionized gas excited by star-formation that we derived are 64\% and 46\% for the lower and higher velocity component, respectively. For the purposes of the calculations of star formation rate (in \S\ref{sec:sfr}) and ionized gas mass (in \S\ref{sec:ion_mass}) we estimate the H$\alpha$ luminosity, coming only from star-formation, by $\mathrm{L(H\alpha)_{SF} = 0.64\,L(H\alpha)_{low} + 0.46\,L(H\alpha)_{high}}$; where $\mathrm{L(H\alpha)_{low}}$ and $\mathrm{L(H\alpha)_{high}}$ are the extinction corrected total H$\alpha$ luminosities in the low and high velocity components, respectively. 

This gives us an extinction-corrected value of $\mathrm{L(H\alpha)_{SF} = 4.99 \pm 0.54 \times 10^{41}\, erg\, s^{-1}}$ for the lower limit to the H$\alpha$ luminosity resulting from star-formation for the Taffy system.

\subsection{Star formation rate estimate from H$\alpha$ luminosity}
\label{sec:sfr}
Using the following relation from \citet{Kennicutt1998} we estimate the star formation rate (SFR) in the entire Taffy system and the bridge region (using the entire region defined as the bridge in Figure \ref{fig:Fig3}), including H$\alpha$ emission from the extragalactic HII region.
\begin{equation}
\mathrm{\psi [M_\odot\, yr^{-1}] = 7.9 \times 10^{-42}  \, L(H\alpha) [erg\, s^{-1}]}
\end{equation}

We obtain $\mathrm{L(H\alpha)_{SF} = 4.99 \pm 0.54 \times 10^{41}\, erg\, s^{-1}}$ and $\mathrm{L(H\alpha)_{SF;bridge} = 1.02 \pm 0.14 \times 10^{41}\, erg\, s^{-1}}$   for the extinction-corrected H$\alpha$ luminosity coming from star-formation for the entire Taffy system and bridge respectively, using the method described previously. This translates to SFRs of 3.94 $\pm$ 1.0 M$_\odot$ yr$^{-1}$ and 0.81 $\pm$ 0.22 M$_\odot$ yr$^{-1}$ respectively for the entire Taffy system and the bridge. These SFRs agree well with our previous estimates derived from UV-FIR SED fitting \citep{Appleton2015} of 3.65~$\pm$~0.03 M$_\odot$ yr$^{-1}$ and 0.69$\pm$0.06 M$_\odot$ yr$^{-1}$ respectively for the total system and the bridge.

Interestingly, recent observations with the Atacama Large Millimeter Array (ALMA) show dense filaments of molecular gas, in the Taffy bridge, with little star-formation in them. These ALMA observations and the overall star-formation properties will be discussed in a future paper (Appleton et al. in preparation).

\subsection{Ionized gas mass}
\label{sec:ion_mass}
The mass of ionized gas in the Taffy system, and in the bridge assuming Case B recombination  \citep{Macchetto1996,Kulkarni2014} is given by :

\vspace{0.2cm}
M$_{ion}$ = 2.33 $\times$ 10$^3$ (L$_{\mathrm{H}\alpha}$/10$^{39}$)(10$^3$/n$_e$) M$_{\odot}$ 

\vspace{0.2cm}
where L$_{\mathrm{H}\alpha}$ is the extinction corrected H$\alpha$ luminosity in units of erg s$^{-1}$, and n$_e$ is the electron density. For the bridge, from the lower limits to the ionized gas fractions from star-formation (see above) we obtained $\mathrm{L(H\alpha)_{SF;bridge} = 1.02 \pm 0.14 \times 10^{41}\, erg\, s^{-1}}$. We also have n$_e$=200 cm$^{-3}$,  based on the ratio of the [SII] lines. This then gives us M$_{ion}$ = 1.19 $\pm$ 0.22 $\times$ 10$^6$ M$_{\odot}$. This calculation is uncertain because the H$\alpha$ emission originates from two different processes -- HII regions and very likely shocks. However, it does show that the ionized gas mass is an insignificant fraction ($\sim$0.2\%) of the total mass of gas in the bridge ($\sim$7 $\times$ 10$^9$ M$_{\odot}$; made up of a mix of HI and H$_2$).  This is in agreement with the very low ionized gas fraction responsible for exciting the [\ion{C}{2}]157.7$\mu$m far-IR cooling line in the bridge \citep{Peterson2018} determined from the upper limit to the detection of [\ion{N}{2}]206$\mu$m in the bridge. For the Taffy system as a whole, using the extinction corrected H$\alpha$ luminosity coming from star-formation, $\mathrm{L(H\alpha)_{SF} = 4.99 \pm 0.54 \times 10^{41}\, erg\, s^{-1}}$, we get M$_{ion}$ = 5.8 $\pm$ 1.0 $\times$ 10$^6$ M$_{\odot}$ for the mass of ionized gas in the entire Taffy system. Again, this is an insignificant fraction ($\sim$0.8\%) of the total gas mass in the Taffy system.

\subsection{Post-starburst populations}
\label{sec:hbeta_ew_results}
We detect H$\beta$ absorption lines within many spaxels on the galaxies. The spectra of post-starburst galaxies are known to contain strong Balmer absorption lines due to their stellar populations being dominated by A type stars. Evidence of a post-starburst population is not uncommon in merging galaxies \citep[e.g.][]{Zabludoff1996, Yang2004, Yang2008}, but attempts to measure the age of the stellar population are difficult, especially when only H$\beta$ is observed \citep[see for e.g.][]{Worthey1997}. Since our spectral coverage did not include other post-starburst indices, we can only provide preliminary results here. Further observations using full UV-optical SED, better absorption line indices, and detailed modeling \citep[e.g.][]{French2016} will be needed to obtain a better estimate for the age for the population which is responsible for the H$\beta$ absorption.

We measured the EW of the H$\beta$ absorption line for each spaxel that contained either the galaxies or the bridge. The EW was measured using,

\begin{equation}
\mathrm{W(H\beta)~[\text{\AA}]} = \frac{\int_{line} f_\lambda d\lambda}{\left<f_{\lambda;cont}\right>}
\end{equation}

where the integral is done over the continuum subtracted absorption line fit and $\left<f_{cont}\right>$ is the average continuum value measured on either side of the H$\beta$ line. LZIFU provides as output the fit to the stellar continuum and the nebular emission lines separately. We use the continuum fit cube to refit a Gaussian absorption line to the region centered on the H$\beta$ absorption. The parameters from this fit then give the area within the absorption line and the average continuum is measured in a band of $\sim$10 spectral elements on both sides of the line.

Figure \ref{fig:halpha_sdss}b shows the measured EW map for the Taffy system. Relatively deep H$\beta$ absorption (W(H$\beta$) $>$ 10) occurs where there is little H$\alpha$ emission. Two regions of high W(H$\beta$) lie in the northern and southern parts of the faint stellar ring that surrounds Taffy-S. Another region, in the south-eastern part of Taffy-N, does extend into regions where there is some star formation and older stellar populations are probably present, with the deepest absorption lying outside of the main star formation disk.

It is generally true that values of W(H$\beta$) of 10 $<$ W(H$\beta$) $<$ 20 implies stellar evolutionary ages for the post-starburst populations of several 100 Myr, and this would imply that there is no connection between the post-starburst population and the current collision between the two galaxies \citep[e.g.][]{Worthey1997}. 
From dynamical arguments it has been argued that the collision between the two Taffy galaxies is quite recent \citep[approximately 25-30 Myrs;][]{Vollmer2012}, and so the fact that the outer ring of Taffy-S shows an old population would lead to an apparent problem, \emph{if} the stellar population of the ring was created in the collision. However, one solution might be that the stellar population of the ring is a stellar density wave containing a much older pre-collisional system which had undergone star formation a long time in the past. This is also the conclusion reached by \citet{Jarrett1999} who argue that the ring consists of stars from an old disk population from the pre-collision disk.

The fact that both galaxies contain evidence of post-starburst activity might imply that the galaxies underwent a high-speed encounter in the more distant past which triggered star formation, but did not lead to immediate merger.  If that is the case, then we are probably currently witnessing the second (probably final) collision before full merger. Broader wavelength coverage in the blue, to detect more absorption lines and better characterize the age of the post-starburst population, will be needed to test this idea.   

\section{Conclusions}
\label{sec:conclusions}
Using visible IFU data, from the VIRUS-P instrument on 2.7m telescope at McDonald Observatory, for the Taffy system, we have shown the results summarized below.

\begin{itemize}
\item{We detect widespread ionized gas within the disks of the Taffy galaxies and the bridge which exhibit very disturbed kinematics, including many regions with double line profiles and emission regions that do not follow regular rotation. Although both galaxies show velocity components that approximate gas rotation around their centers, both galaxies also show peculiar motions, often associated with gas which extends into the bridge between them. The gas associated with Taffy-N (UGC~12915) contains a major kinematic component that does not take part in regular rotation, but exhibits a high velocity dispersion and forms a narrow western bridge to Taffy-S (UGC~12914). Taffy-S, although showing the kinematics of a likely tidally-warped but regular (counter-)rotating disk, also contains a peculiar kinematic component that is associated with a second  ionized gas bridge. This eastern bridge component, which extends from Taffy-N, through the region containing the extragalactic HII region and eventually connecting with the Taffy-S, is more closely associated with the molecular bridge seen previously to extend between the galaxies. On the other hand, the western bridge, which is much fainter, appears to be kinematically linked with the western extension of Taffy-N (and the western part of Taffy-S) and shows a mix of HII-region and composite/LINER excitation unlike the main disk of Taffy-N which is largely consistent with HII-region excitation.
}

\item{An analysis of the excitation of the ionized gas through diagnostic line ratios shows that a significant fraction of the emission shows a mix of HII-region and LINER-type emission, especially in the areas where we can clearly distinguish two velocity components. The LINER-type emission is especially dominant in the high velocity component in the east-bridge region, but also over significant portions of both galaxies. We observe emission line ratios in the high-velocity component, for the east bridge and the low-velocity component of the west bridge, that are consistent with the gas being excited by shocks with velocities of $\sim$175-200 km/s, and a range of pre-shock densities. Such evidence for shocks permeating clouds of varying density is consistent with previous observations of the bridge by {\it Spitzer} and {\it Herschel}, where strong mid- and far-IR cooling lines are detected from warm molecular and diffuse atomic gas heated by turbulence. While we cannot rule out a contribution to the diffuse ionized emission from Lyman-continuum photons leaking from young HII regions embedded in the disks of both galaxies, the weight of evidence, from previous multi-wavelength observations of the bridge, suggests that shocks are a more likely explanation for the LINER-type emission line ratios seen in the bridge and in parts of the disks of both galaxies. Given the violence of the collision based on previous numerical models of  head-on collisions which suggest that the bridge is still in a highly disturbed state, such shocks may not be unexpected. 
}

\item{Strong Balmer absorption lines ($10 < \mathrm{W(H\beta)} [\text{\AA}] < 15$) are observed in parts of the ring associated with Taffy-S (UGC~12914), as well as the south-east portion of the edge-on disk of Taffy-N (UGC~12915). The absorption lines are strongest in regions where the ionized and molecular gas distributions are weak, suggesting that parts of the Taffy system have experienced a burst of star formation in the past,  perhaps from a previous close passage of the two galaxies. If so, it is possible that the current collision may be a second, more dissipative collision, that will likely lead to merger in the near future.   Further observations, with a broader blue wavelength coverage than the current observations, will be necessary to better determine the age of the post-starburst populations in both galaxies.
}

\item{Although there has been only weak evidence in the past for the nucleus of Taffy-S (UGC 12914) containing a low-luminosity AGN from X-ray properties \citep{Appleton2015}, we detect line widths in the nucleus as large as 320 \kms.  These line-widths are typical of narrow-line Seyfert galaxies. The broad lines appear to extend over 6-10 arcsecs along the minor axis.
The excitation properties of the nuclear gas are consistent with LINER emission, but without higher spatial and velocity resolution data we cannot determine whether the Taffy-S nucleus hosts a weak shocked highly confined outflow from a nuclear starburst, or is excited by UV radiation from a LLAGN.
}

\item{We provide evidence, supporting much previous work, that the Taffy system has atypically low SFRs for a system having recently undergone a recent major-merger. We find SFRs of 3.94 $\pm$ 1.0 M$_\odot$ yr$^{-1}$ and 0.81 $\pm$ 0.22 M$_\odot$ yr$^{-1}$ in the entire Taffy system and the bridge (including the extragalactic HII region), respectively. Low star formation rates in this recent post-collisional remnant may result from the highly disturbed nature of the gas in the galaxies and bridge.}
\end{itemize}

\acknowledgments
BAJ would like to thank the Visiting Graduate Fellowship Program at IPAC/Caltech for 6 months support towards work performed for this paper. BAJ is also grateful to Drs.\ Rogier Windhorst and Rolf Jansen for helpful discussions on work done in this paper. The authors thank the anonymous referee for their helpful review and suggestions. This research has made use of NASA's Astrophysics Data System. This research has made use of the Python programming language along with the Numpy, Scipy, and Matplotlib packages. This research has also made use of Astropy, a community-developed core Python package for Astronomy \citep{astropy2018}.

\clearpage  

\LongTables
\begin{deluxetable*}{c p{4cm} p{9cm}}
\tablecaption{LZIFU parameters supplied to the configuration file. \label{tab:lzifu_params}}
\label{{tab:lzifu_params}}
\tabletypesize{\scriptsize}
\tablehead{
\colhead{Parameter name} & \colhead{Value} & \colhead{Description}
}
\startdata
\texttt{only\_1side} & 0 & 0: 2-sided data. 1: 1-sided data \\
\texttt{z} & 0.0145 & Redshift \\
\texttt{fit\_ran} & [4700,6855] & Fitting range \\
\hline
\multicolumn{3}{c}{Continuum fitting with PPXF} \\
\hline
\texttt{mask\_width}             & 12 &  Full width to mask around emission lines defined in lzifu\_linelist.pro \\
\texttt{cont\_vel\_sig\_guess}     & [0., 50.] & Starting guess of delV and vel\_dispersion of continuum (in km/s) \\
\texttt{cont\_ebv\_guess}         & 0.1  & Starting guess of ebv \\
\hline
\multicolumn{3}{c}{Emission fitting with MPFIT} \\
\hline
\texttt{fit\_dlambda}    & 22. & Full width around line centers to be fitted. (A)  \\
\texttt{ncomp}          & 2 & Number of component.  \\
\texttt{line\_sig\_guess} & 70. & Initial guess of velocity dispersion for emission line (km/s) \\
\texttt{vdisp\_ran}      & [-50,500.]   & Velocity dispersion constraints in km/s.  \\
\texttt{vel\_ran}        & [-600.,+600.] & Velocity contriants in km/s. 0 is systemic velocity from set.z \\
\hline
\multicolumn{3}{c}{Variation in initial guess\footnote{LZIFU explores all possible combinations of initial guesses of 1$^{st}$ and 2$^{nd}$ components.}} \\
\hline
\texttt{comp\_2\_damp}    & [0.6] &  Initial guess for amplitude of 2$^{nd}$ component as fraction of 1$^{st}$ component amplitude \\
\texttt{comp\_2\_dvel}    & [-150,-50,+50,+150] & Initial guess range for velocity of 2$^{nd}$ component; given as difference between velocities of 1$^{st}$ and 2$^{nd}$ components \\
\texttt{comp\_2\_dvdisp}  & [+20] & Initial guess for velocity dispersion of 2$^{nd}$ component \\
\enddata
\end{deluxetable*}

\begin{thebibliography}{}
\bibitem[Adams et al.(2011)]{Adams2011} Adams, J. J., Blanc, G. A., Hill, G. J., et al.\ 2011, ApJS, 192, 5
\bibitem[Alatalo et al.(2015)]{Alatalo2015} Alatalo, K., Appleton, P. N., Lisenfeld, U., et al.\ 2015, ApJ, 812, 117
\bibitem[Allen et al.(2008)]{Allen2008} Allen, M. G., Groves, B. A., Dopita, M. A., Sutherland, R. S., \& Kewley, L. J.\ 2008, \apjs, 178, 20-55 
\bibitem[Appleton et al.(2006)]{Appleton2006} Appleton, P. N., Xu, K. C., Reach, W., et al.\ 2006, ApJ, 639, 51
\bibitem[Appleton et al.(2015)]{Appleton2015} Appleton, P. N., Lanz, L., Bitsakis, T., et al.\ 2015, ApJ, 812, 118
\bibitem[Appleton et al.(2017)]{Appleton2017} Appleton, P. N., Guillard, P., Togi, A., et al.\ 2017, ApJ, 836, 76
\bibitem[Armus et al.(1987)]{Armus1987} Armus, L., Heckman, T., Miley, G., 1987, AJ, 94, 831
\bibitem[Armus et al.(2009)]{Armus2009} Armus, L., Mazzarella, J. M., Evans, A. S., et al.\ 2009, PASP, 121, 559
\bibitem[Astropy Collaboration,(2018)]{astropy2018} The Astropy Collaboration, arxiv:1801.02634
\bibitem[Baldwin et al.(1981)]{Baldwin1981} Baldwin, J. A., Phillips, M. M., \& Terlevich, R., 1981, PASP, 93, 5
\bibitem[Blanc et al.(2010)]{Blanc2010} Blanc, G. A., Gebhardt, K., Heiderman, A., et al.\ 2010, New Horizons in Astronomy: Frank N. Bash Symposium 2009, 432, 180 
\bibitem[Blanc et al.(2013)]{Blanc2013} Blanc, G. A., Weinzirl, T., Song, M., et al.\ 2013, \aj, 145, 138 
\bibitem[Braine et al.(2003)]{Braine2003} Braine, J., Davoust, E., Zhu, M., et al.\ 2003, A\&A, 408, 13
\bibitem[Braine et al.(2004)]{Braine2004} Braine, J., Lisenfeld, U., Duc, P.-A., et al.\ 2004, A\&A, 418, 419
\bibitem[Brinchmann et al.(2004)]{Brinchmann2004} Brinchmann, J., Charlot, S., White, S. D. M., et al.\ 2004, MNRAS, 351, 1151
\bibitem[Bushouse(1986)]{Bushouse1986} Bushouse, H. A., 1986, AJ, 91, 255
\bibitem[Bushouse(1987)]{Bushouse1987} Bushouse, H. A., 1987, ApJ, 320, 49
\bibitem[Calzetti et al.(2000)]{Calzetti2000} Calzetti, D., Armus, L., Bohlin, R. C., et al.\ 2000, \apj, 533, 682
\bibitem[Calzetti (2001)]{Calzetti2001} Calzetti, D., 2001, PASP, 113, 1449
\bibitem[Capellari \& Emsellem(2004)]{Capellari2004} Capellari, M. and Emsellem E., 2004, PASP, 116, 138
\bibitem[Cluver et al.(2010)]{Cluver2010} Cluver, M. E., Appleton, P. N., Boulanger, F., et al.\ 2010, ApJ, 710, 248
\bibitem[Cluver et al.(2013)]{Cluver2013} Cluver, M. E., Appleton, P. N., Ogle, P., et al.\ 2013, \apj, 765, 93
\bibitem[Condon et al.(1993)]{Condon1993} Condon, J. J., Helou, G., Sanders, D. B., \& Soifer, B. T., 1993, AJ, 105, 5
\bibitem[Daddi et al.(2007)]{Daddi2007} Daddi, E., Dickinson, M., Morrison, G., et al.\ 2007, ApJ, 670, 156
\bibitem[Daddi et al.(2010)]{Daddi2010} Daddi, E., Elbaz, D., Walter, F., et al.\ 2010, ApJ, 714L, 118
\bibitem[Driver et al.(2007)]{Driver2007} Driver, S. P., Popescu, C. C., Tuffs, R. J., et al.\ 2007, \mnras, 379, 1022
\bibitem[Elbaz et al.(2002)]{Elbaz2002} Elbaz, D., Cesarsky, C. J., Chanial, P., et al.\ 2002, A\&A, 384, 848
\bibitem[Elbaz et al.(2007)]{Elbaz2007} Elbaz, D., Daddi, E., Le Borgne, D., et al.\ 2007, A\&A, 468, 33
\bibitem[French et al.(2016)]{French2016} French, K. D., Arcavi, I., \& Zabludoff, A.\ 2016, ApJL, 818, 21
\bibitem[Gao et al.(2003)]{Gao2003} Gao, Y., Zhu, M., \& Seaquist, E. R. 2003, AJ, 126, 2171
\bibitem[Genzel et al.(2010)]{Genzel2010} Genzel, R., Tacconi, L. J., Gracia-Carpio, J., et al.\ 2010, MNRAS, 407, 2091
\bibitem[G{\"u}ver \& {\"O}zel(2009)]{Guver2009} G{\"u}ver, T., \& {\"O}zel, F.\ 2009, \mnras, 400, 2050 
\bibitem[Guillard et al.(2009)]{Guillard2009} Guillard, P., Boulanger, F., Pineau Des ForÃªts, G., \& Appleton, P. N. 2009, A\&A, 502, 515
\bibitem[Hampton et al.(2017)]{Hampton2017} Hampton, E.~J., Groves, B., Medling, A., et al.\ 2017, Astronomical Data Analysis Software and Systems XXV, 512, 221
\bibitem[Hill et al.(2008)]{Hill2008} Hill, G. J.; MacQueen, P. J.; Smith, M. P. et al.\ 2008, SPIE, 7014, 70
\bibitem[Ho et al.(2016)]{Ho2016} Ho, I. T.; Medling, A. M.; Groves, B. et al.\ 2016, Ap\&SS, 361, 280
\bibitem[Jarrett et al.(1999)]{Jarrett1999} Jarrett, T. H., Helou, G., Van Buren, D., Valjavec, E., \& Condon, J. J.\ 1999, \aj, 118, 2132
\bibitem[Joseph \& Wright(1985)]{Joseph1985} Joseph, R. D. \& Wright, G. S., 1985, MNRAS, 214, 87
\bibitem[Kauffmann et al.(2003)]{Kauffmann2003} Kauffmann G., Heckman, T. M., Tremonti, C., et al.\ 2003, MNRAS, 346, 1055
\bibitem[Kennicutt(1998)]{Kennicutt1998} Kennicutt, R. C. Jr., 1998, ARA\&A, 36, 189
\bibitem[Kewley et al.(2001)]{Kewley2001} Kewley, L. J., Heisler, C. A., Dopita, M. A., \& Lumsden, S.\ 2001, \apjs, 132, 37 
\bibitem[Kewley et al.(2006)]{Kewley2006} Kewley, L. J., Groves, B., Kauffmann, G., \& Heckman, T., 2006, MNRAS, 372, 961
\bibitem[Kulkarni et al.(2014)]{Kulkarni2014} Kulkarni, S., Sahu, D. K., Chaware, L., Chakradhari, N. K., \& Pandey, S. K.\ 2014, New Astronomy, 30, 51 
\bibitem[Komugi et al.(2012)]{Komugi2012} Komugi, S., Tateuchi, K., Motohara, K., et al.\ 2012, \apj, 757, 138
\bibitem[Levesque \& Leitherer(2013)]{Levesque2013} Levesque, E.~M., \& Leitherer, C.\ 2013, \apj, 779, 170
\bibitem[Lisenfeld \& V\"olk(2010)]{Lisenfeld2010} Lisenfeld, U. \& V\"olk, H. J. 2010, A\&A, 524, 27
\bibitem[Lisenfeld et al.(2017)]{Lisenfeld2017} Lisenfeld, U., Alatalo, K., Zucker, C., et al.\ 2017, \aap, 607, A110 
\bibitem[Macchetto et al.(1996)]{Macchetto1996} Macchetto, F., Pastoriza, M., Caon, N., et al.\ 1996, \aaps, 120, 463 
\bibitem[Markwardt(2009)]{Markwardt2009} Markwardt, C. B., 2009, ASPC, 411, 251
\bibitem[Peterson et al.(2012)]{Peterson2012} Peterson, B. W., Appleton, P. N., Helou, G., et al.\ 2012, ApJ, 751, 11
\bibitem[Peterson et al.(2018)]{Peterson2018} Peterson, B. W., Appleton, P. N., Bitsakis, T., et al.\ 2018, ApJ, 855, 141 
\bibitem[Rich et al.(2011)]{Rich2011} Rich, J. A., Kewley, L. J., Dopita, M. A. 2011, ApJ, 734, 87
\bibitem[Rich et al.(2014)]{Rich2014} Rich, J. A., Kewley, L. J., Dopita, M. A. 2014, ApJ, 781, 12
\bibitem[Rich et al.(2015)]{Rich2015} Rich, J. A., Kewley, L. J., Dopita, M. A. 2015, ApJS, 221, 28
\bibitem[Sanders et al.(1988a)]{Sanders1988a} Sanders, D. B., Soifer, B. T., Elias, J. H., et al.\ 1988, ApJ, 325, 74
\bibitem[Sanders et al.(1988b)]{Sanders1988b} Sanders, D. B., Soifer, B. T., Elias, J. H., et al.\ 1988, ApJ, 328, 35
\bibitem[Sanders \& Mirabel(1996)]{Sanders1996} Sanders, D. B. \& Mirabel, I. F. 1996, ARA\&A, 34, 749
\bibitem[Sanders et al.(2003)]{Sanders2003} Sanders, D. B., Mazzarella, J. M., Kim, D.-C., Surace, J. A., \& Soifer, B. T.\ 2003, AJ, 126, 1607 
\bibitem[Santini et al.(2009)]{Santini2009} Santini, P., Fontana, A., Grazian, A., et al.\ 2009, A\&A, 504, 751
\bibitem[Soifer et al.(1987)]{Soifer1987} Soifer, B. T., Sanders, D. B., Madore, B. F., et al.\ 1987, ApJ, 320, 238
\bibitem[Soifer \& Neugebauer(1991)]{Soifer1991} Soifer, B. T. \& Neugebauer, G., 1991, AJ, 101, 354
\bibitem[Stark et al.(2009)]{Stark2009} Stark, Daniel P., Ellis, Richard S., Bunker, Andrew, et al.\ 2009, ApJ, 697, 1493
\bibitem[Struck(1997)]{Struck1997} Struck, C.\ 1997, ApJS, 113, 269 
\bibitem[Tacconi et al.(2010)]{Tacconi2010} Tacconi, L. J., Genzel, R., Neri, R., et al.\ 2010, Nature, 463, 781
\bibitem[Unterborn \& Ryden(2008)]{Unterborn2008} Unterborn, C. T. \& Ryden, B. S.\ 2008, \apj, 687, 976
\bibitem[van-Dokkum(2001)]{vanDokkum2001} van-Dokkum, P., 2001, PASP, 113, 1420
\bibitem[Veilleux \& Osterbrock(1987)]{Veilleux1987} Veilleux, S., \& Osterbrock, D. E.\ 1987, \apjs, 63, 295
\bibitem[Vollmer et al.(2012)]{Vollmer2012} Vollmer, B., Braine, J., \& Soida, M. 2012, A\&A, 547, 39
\bibitem[Worthey \& Ottaviani(1997)]{Worthey1997} Worthey, G.\ \& Ottaviani, D.\ L.\ 1997, ApJS, 111, 377
\bibitem[Yang et al.(2004)]{Yang2004} Yang, Y., Zabludoff, A. I., Zaritsky, D.\ et al.\ 2004, ApJ, 607, 258
\bibitem[Yang et al.(2008)]{Yang2008} Yang, Y., Zabludoff, A. I., Zaritsky, D., \& Mihos, C.\ J.\ 2008, ApJ, 688, 945
\bibitem[York et al.(2000)]{York2000} York, D. G.; Adelman, J.; Anderson, J. E., Jr. et al.\ 2000, AJ, 120, 1579
\bibitem[Weilbacher et al.(2018)]{Weilbacher2018} Weilbacher, P.~M., Monreal-Ibero, A., Verhamme, A., et al.\ 2018, A\&A, 611, 95
\bibitem[Zabludoff et al.(1996)]{Zabludoff1996} Zabludoff, A. I., Zaritsky, D., Lin, H.\ et al.\ 1996, ApJ, 466, 104
\bibitem[Zhu et al.(2007)]{Zhu2007} Zhu, M., Gao, Y., Seaquist, E. R. et al.\ 2007, AJ, 134, 118
\end{thebibliography}
\end{document}